\documentclass{iopart}

\usepackage{graphicx}
\usepackage{iopams}
\usepackage{setstack}

\newcommand{\figwidth}{0.6\columnwidth}

\begin{document}

\title{Influence of non-magnetic impurities on hole doped two-leg $Cu$-$O$ Hubbard ladders}
\author{P. Chudzinski$^{1,2}$, M Gabay$^2$ and T Giamarchi$^1$}
\address{$^1$ DPMC-MaNEP, University of Geneva, 24 Quai
Ernest-Ansermet CH-1211 Geneva, Switzerland}
\address{$^2$ Laboratoire de Physique des Solides, Bat. 510,
Universit\'e Paris-Sud 11, Centre d'Orsay, 91405 Orsay Cedex,
France}
\begin{abstract}

We study the influence of non magnetic impurities on the phase
diagram of doped two-leg Hubbard $Cu$-$O$ ladders. In the
absence of impurities this system posseses d-wave
superconducting states and orbital current states depending on
the doping. A single, strong, scatterer modifies its
environment locally and this effect is assessed using a
renormalization group analysis. At high doping, disorder causes
intraband instabilities and at low doping it promotes interband
instabilities. In the former case, we extend the boundary
conformal field theory method -- developed in the context of
single chains -- to handle the ladder problem, and we find
exact closed-form analytical expressions for the correlation
functions. This allows us to compute experimentally measurable
local quantities such as the nuclear magnetic resonance line
broadenings and scanning tunnelling microscope profiles. We
also discuss the low doping regime where Kondo physics is at
play, making qualitative predictions about its nature. Insight
into collective effects is also given in the many weak
impurities case, based on an RG approach. In this regime, one
sees the interplay between interactions and disorder. We
emphasize the influence of the $O$ atoms on disorder effects
both for the single- and for the many- defect situations.

\end{abstract}
\maketitle

\section{Introduction}

Despite intensive efforts, the interplay between interactions
and superconductivity  in low dimensional systems is still
poorly understood. An especially important question,
considering the case of high Tc compounds, is whether
interactions are responsible for the unusually high
superconducting temperatures in these materials. In addition to
this issue, there is a considerable debate on the appropriate
choice of a minimal model needed to describe these compounds. A
remarkable situation when these questions can be studied in a
controlled fashion is the one dimensional limit of ladder
materials \cite{dagotto_ladder_review}. Indeed, in one
dimension, powerful theoretical tools \cite{giamarchi_book_1d}
allow a detailed analysis of interactions effects
\cite{varma_2bands,nersesyan_2ch,fabrizio_2ch_rg,schulz_moriond,kuroki_2chain_phases,ledermann_2band_spinless,balents_2ch,lin_Nchains}.
It was shown, for a single band model, that ladders exhibit a
form of d-wave superconductivity driven by purely repulsive
interactions, and this result is suggestive that a similar
physics might be at play in their higher dimensional
counterpart.
In addition to their interest in connection with the cuprates,
ladders have spawned their own field, thanks to the existence
of several experimental realizations \cite{piskunov_ladder_nmr,
piskunov04_sr14cu24o41_nmr,fujiwara03_ladder_supra,
imai_NMR_doped_2ladder,kumagai_NMR_2ladder}.

In addition to the superconducting phase,
many questions touch upon the normal phase in these
superconducting systems and in particular the topic of the
so called pseudogap phase. Various explanations have
been put forward, ranging from preformed superconducting pairs
 \cite{kotliar_liu_dwave_slavebosons} to the existence of
orbital currents with \cite{chakravarty_ddw_pseudogap} or
without \cite{cvarma_3band_model,CVarma_orbitalcurrents} broken
translational symmetry. Recent neutrons experiments
 \cite{BFauque_neutrons,li_orbital_mercury} indicating the
presence of magnetic moments, compatible with translationnally
invariant pattern of currents, and Kerr effect measurements
 \cite{Xia_optOAF} showing evidence of time-reversal symmetry
breaking have generated a large interest in this topic.
Although current (flux) phases have been first proposed for the
single-band Hubbard model
 \cite{affleck_marston,kotliar_liu_dwave_slavebosons,lederer_fluxphases,zhang_flux_ref},
they have been found unstable in slave bosons and numerical
calculations. Ladders have also provided a controlled
environment to study this important issue. It was found that
somewhat special interactions, more complex than local ones,
are needed to stabilize them
 \cite{orignac_2chain_long,schollwock_CDW+current}. The
resulting phases break the translational symmetry of the
lattice, leading to a staggered flux pattern. Similar staggered
patterns were advocated as a potential explanation of the
pseudogap phase \cite{chakravarty_ddw_pseudogap} (DDW phase).

In order to find new phases it thus seemed necessary to go
beyond the single band model
\cite{cvarma_3band_model,CVarma_orbitalcurrents}. For ladders,
this was done both at half filling \cite{lee_marston_CuO} when
the system is insulator and more recently
 \cite{chudzinski_ladder_rapid,chudzinski_ladder_long} using a
renormalization procedure, for arbitrary doping. Quite
remarkably for finite doping, a new massless phase, absent in
the case of a single band Hubbard ladder and able to sustain
orbital currents was found. A detailed investigation of the
properties of this new phase, and more generally of a three
band model for a ladder system is thus a challenging and
interesting question.

One way to access the properties of such a phase is to probe
its response to the presence of impurities. Indeed, when
defects are present, translational symmetry is broken and
momentum is not a good quantum number anymore. Wavefunctions
with different momenta mix and, at low temperature, the static
($\omega =0$) response contains contributions from large
momentum states. Investigations of correlation effects using
impurities was largely done for \emph{insulating} ladders.
Nuclear magnetic resonance (NMR) experiments performed by
Takigawa et. al. \cite{Takigawa_Zn_chains} showed that
non-magnetic impurities broaden the NMR lines, which then
display a distinctive temperature dependence. This finding
confirmed earlier theoretical predictions \cite{Eggert_gen1}.
Subsequently, various families of low dimensional models with
different characteristics (dimerization, zig-zag) were
investigated theoretically \cite{Laukamp_DMRG_imp}. In the
vicinity of an impurity, and in the ground state, two distinct
generic behaviors were found 1) a magnetic state is
exponentially localized near the defect 2) a delocalized
antiferromagnetic (AF) cloud surrounds the defect (for
temperatures on the order of 100K, the polarization extends
over tens of lattice spacings). The former case corresponds to
systems in which magnetic degrees of freedom are gapped (e.g
the Haldane spin $1$ chain \cite{haldane_gap}) while the latter
case corresponds to an ungapped situation (e.g the spin $1/2$
chain or the pseudo-gap state of two-dimensional systems). Both
behaviors were found experimentally. For instance, on the basis
of an NMR study on ladders, the authors of
Ref.~\cite{Fujiwara_NMR_Zn} claimed that the temperature
dependence of the line broadening could be explained by the
former (gapped) case for very small concentrations of defects
and by the latter (ungapped) case for larger concentrations.
 Similar results and conclusions were reported in the work of
Ref.~\cite{Ohsugi_NMR_Zn}, on the basis of NMR and NQR
measurements. From the amplitude of the broadened part, both
sets of authors concluded that the AF cloud has to extend over
surprisingly many unit cells.

On the theoretical side, Laukamp et al. \cite{Laukamp_DMRG_imp}
performed numerical studies of 1D spin models, including spin
ladders. They observed that, quite generally, the magnetic
response of the system contains a large staggered part near an
isolated defect. Further computational work was done by Lauchli
et al. \cite{Lauchli_Li_imp}, who studied the nature of the
states around a Li atom impurity substituting a copper site in
a single orbital $Cu$ ladder. They found a triplet bound state
but suggested that free spinons could be obtained when several
defects are present or with other types of impurities. The
former case is suggestive of the overscreened free spin
scenario which is usually discussed in the context of the
density matrix renormalization group (DMRG) analysis of two-leg
spin ladders \cite{Laukamp_DMRG_imp}. The latter case signals
that, even with a vanishingly small density of defects, spinons
will gain kinetic energy and delocalize rather than  form a
local bound state inside the energy gap, close to the impurity.
These two different regimes were also observed  in studies of
the quasiclassical sine-Gordon model \cite{Nakamura_soliton}.
The bound  spin state limit was investigated analytically,
using a Majorana fermion technique \cite{Lecheminant_Majorana}.
The authors of that study stated the conditions under which
this regime might exist and computed the additional
contribution to the susceptibilities stemming from the bound
state. The properties of soliton states at the boundary (i.e
next to the defect) -- including magnetic susceptibilities --
were discussed in detail in a series of papers by Essler
\cite{schuricht_gapped_strong_V, Schuricht_boundary_sG}.

In this work, we consider the problem of impurities in
\emph{doped} two-leg $Cu$-$O$ Hubbard ladders for which both
the superconducting phases and the novel orbital current phases
are present. In the defect-free case, the phase diagram as a
function of carrier doping was established in Refs.~
\cite{chudzinski_ladder_rapid,chudzinski_ladder_long}. Such a
system is characterized by two energy bands controlling the
fluctuations of (two) charge and (two) spin modes near the
Fermi energy.
Three different regimes were
found: 1) strong gaps in the spin modes and in one of the
charge modes at low doping; 2) a Luttinger liquid (LL)
(gapless) phase at intermediate doping; 3) gapless charge
modes, no gap for one of the spin mode and a small gap for the
other at higher doping. Because the most significant
perturbation caused by impurities is the mixing of the two
bands we can expect this effect to be the largest in the last
two doping ranges where interactions promote band separation in
the absence of disorder. By contrast, in the ground state of
the clean system at low doping, the two bands strongly mix so
that the perturbation is expected to be less ``dangerous'' and
the physics of the lightly doped ladder should be qualitatively
similar to that of the disordered (undoped) spin ladder. For
this reason, we will be mostly concerned in this paper with the
regime of intermediate and large carrier content. For large
dopings, in the absence of disorder, the size of the gap in the
spin sector is small, so that, for experimentally relevant
temperatures, it does not play a role, and the state is
effectively described by a LL. Hence, from this point on, we
focus on the role of disorder in the LL phase of moderately and
highly doped $Cu$-$O$ ladders. We will consider two extreme
situations, for impurities, namely the case of a single
impurity producing a strong scattering potential on the one
hand and the case of a finite concentration of defects
producing each a weak scattering potential, on the other.

Our paper is organized as follows: in Sect.~\ref{sec:model}, we
present the model which describes the physics of a doped
two-leg $Cu$-$O$ Hubbard ladder. In the continuum limit the
quadratic part of the Hamiltonian is diagonal in a particular
basis where it takes the simple form of a sum of LL modes. The
relevant notation used in the course of this paper is
introduced at that stage. We recall the characteristics of the
phase diagram in the absence of disorder. The hamiltonian
modelling the scattering by a single, strong impurity is then
expressed in terms of bosonic phase fields.

In Sec.~\ref{CFT}, we generalize the boundary conformal field
theory (BCFT) methods -- which have been successfully used
in the context of chains -- to handle the two-leg ladder problem.
The possible fixed points of the RG flows are given in the
presence of disorder, and their stability is discussed. Then,
the generalized boundary conditions for the phase fields are
derived, at the open boundary condition (OBC) fixed point.

Sec.~\ref{resultats} presents our main results for the single
impurity case. Only local properties of the system are
affected, and, when we are considering LL phases at
intermediate and high dopings, the bulk parameters $K_{i}$ (the
compressibilities) remain unchanged. We generalize the
procedure developed by Eggert and Affleck \cite{Eggert_gen1,
Eggert2} to treat the two-band case. Green functions and
alternating magnetizations in the vicinity of the impurity are
computed.  $\chi_{alt}$, the alternating part of the (local)
susceptibility is predicted to cause NMR Knight shift
broadenings $\Delta$K, and the changes in the imaginary part of
the Green function ought to be detectable in scanning
tunnelling microscopy (STM) measurements of the local spectral
weights (LSW). We study their temperature dependence in order
to provide experimentally testable predictions. Possible
differences stemming from the location of the impurity within
the unit cell or outside of the ladder are discussed.

In Sec.~\ref{discussion}, for the sake of completeness, we
consider various limiting situations. First we discuss the nature
of the states in the low doping phase. Next, we examine the
rigidity of intermediate, gapless phase, and consider a putative
instability of the bulk phase in the vicinity of the impurity. The
competition between disorder and interactions is discussed in
terms of RG arguments. Finally, in the presence of many impurities
acting collectively, we analyse the RG flow leading to Anderson
localization.

We conclude, with an experimental perspective on the use of
non-magnetic impurities as a probe of correlated ground states
in such ladder systems.

\section{The model}\label{sec:model}

We consider a two-leg $Cu$-$O$ ladder with an impurity.
The Hamiltonian of this system contains three parts:
the kinetic energy of electrons moving on the lattice
$H_{T}$,  electron interactions $H_{int}$ and an impurity term
$H_{imp}$
\begin{equation}\label{hubb}
    H=H_{T}+H_{int}+H_{imp}
\end{equation}
The physics of the first two terms is known
 \cite{chudzinski_ladder_rapid,chudzinski_ladder_long}. In the
first part of this section, we briefly recall its main
characteritics, and we also introduce relevant notations. In the
second part, we express the impurity Hamiltonian using the
standard bosonization description of one dimensional systems
 \cite{giamarchi_book_1d}.

\subsection{The clean ladder}\label{cleanlad}

Including two $Cu$ and five $O$ atoms in the unit cell, the
tight-binding kinetic energy part $H_{T}$ reads
\begin{eqnarray}
 \fl H_{T} = \sum_{j\sigma}( \sum_{m\in Cu}\epsilon_{Cu}
    n_{mj\sigma}+ \sum_{m\in O}\epsilon_{O}
    n_{mj\sigma}- \sum_{m\in Cu} t\cdot
    [a_{mj\sigma}^{\dag}(b_{mj\sigma}+b_{mj-1,\sigma})+h.c.] \nonumber\\
     - \sum_{m\in Cu} t_{\bot}\cdot
    [a_{mj\sigma}^{\dag}(b_{m+1,j\sigma}+b_{m-1,j\sigma})+h.c.]) \label{hubbT} \\
     -\sum_{m=\in O(leg)} t_{pp}\cdot
    [b_{mj\sigma}^{\dag}(b_{m+1,j\sigma}+b_{m-1,j\sigma}+b_{m+1,j-1\sigma}+b_{m-1,j-1\sigma})+h.c.])\nonumber
\end{eqnarray}
where $a_{mj\sigma}(b_{mj\sigma})$ is the creation operator of
holes with spin $\sigma$ on a copper (oxygen) site (\emph{j} is a
site along chain and \emph{m} labels the atoms in each cell);
$n_{mj\sigma}^{Cu}=a_{mj\sigma}^{\dag}\cdot a_{mj\sigma}$
We use hole notation such that $t$,
$t_{\bot}$,$t_{pp}$ are all positive.
$\epsilon=\epsilon_{O}-\epsilon_{Cu}$ is the difference between
the oxygen and copper on-site energies.

Eigenvalues and eigenvectors of the non-interacting part are
simply obtained through a Fourier transform of $H_{T}$. If
$\epsilon$ is of the same order as- or larger than- the hopping
\emph{t}, one may neglect the non-bonding and anti-bonding high
energy states which have mostly a p-type character, and this
reduces the model to two low lying bands crossing the Fermi
energy. They are denoted $o$ (symmetric under the exchange of
the two legs) and $\pi$ (antisymmetric under the exchange of the
two legs). With $\alpha$ ($=0,\pi$) and
$\sigma$ denoting the band and the spin index respectively, the Hamiltonian reads
\begin{equation}\label{HTdiag}
    H_{T}=\sum_{k\alpha\sigma}e_{\alpha}(k)n_{k\alpha\sigma}
\end{equation}
where
\begin{equation}\label{bazy1}
    a_{mk\sigma}= \sum_{\alpha} \lambda_{m\alpha} a_{\alpha k\sigma}
\end{equation}
and $e_{\alpha}(k)$ and $\lambda_{m\alpha}$ are the eigenvalues
and components of the eigenvectors of the Hamiltonian matrix
(see Ref.~\cite{chudzinski_ladder_long}).

In the low energy limit one may linearize the dispersion
relation in the vicinity of the Fermi energy:
\begin{equation}\label{contin}
    H_{T}=\sum_{|q|<Q} \sum_{r\alpha\sigma}r q V_{F\alpha} a^{\dag}_{\alpha r
    q\sigma} a_{\alpha r q\sigma}
\end{equation}
and it is easy to bosonize this free fermion theory
 \cite{giamarchi_book_1d,voit_bosonization_revue}. Two (charge
and spin) boson phase fields denoted by $\phi_{\nu}(x)$ are
introduced for each fermion specie. $\nu$ denotes both spin and
band indexes, $x$ is the spatial coordinate along the ladder.
Fermionic operators are written as
$\psi_{\nu}\simeq\eta_{\nu}\exp(\imath
(\phi_{\nu}+\theta_{\nu}))$ where $\eta_{\nu}$ are the Klein
factors which satisfy the required anticommutation relations
for fermions. These $\eta_{\nu}$ do not contain any spatial
dependence and they commute with the Hamiltonian operator. They
only influence the form of the order operators in bosonic
language (through contributions $\eta_{\nu}\eta_{\nu'}$) and
the signs of the non-linear couplings via $\Gamma$ coefficients
(eigenvalues of
$\eta_{\nu}\eta_{\nu'}\eta_{\nu''}\eta_{\nu'''}$ operators).
$\Gamma^{2}=1$ and in this paper we choose the convention
$\Gamma=+1$, as in Ref.~\cite{tsuchiizu_2leg_firstorder}.

We also introduce the phase field $\theta_{\nu}(x)$; its
spatial derivative $\Pi_{\nu}(x)$ is canonically conjugated with
$\phi_{\nu}(x)$. Now the Hamiltonian may be rewritten using the above
phase fields. The kinetic part and those pieces in $H_{int}$ which
can be expressed as density-density terms give rise to the following
form
\begin{equation}\label{eq:Hbozon}
    H_{0}= \sum_{\nu} \int \frac{dx}{2\pi}[(u_{\nu}K_{\nu})(\pi \Pi_{\nu})^{2}+(\frac{u_{\nu}}{K_{\nu}})(\partial_{x} \phi_{\nu})^{2}]
\end{equation}
For the non-interacting system one has $K_{\nu}=1$ for all modes;
the diagonal density basis is then the bonding/antibonding one
$B_{o\pi}$ (the momentum $k_{\perp}$ associated with the rungs is
either 0 or $\pi$). The other basis which is commonly used in the
literature is the total/transverse one, $B_{+-}$. It is
related to $B_{o\pi}$ in such a way that:
\begin{equation}\label{eq:basis}
    \phi_{\nu+(-)}=\frac{\phi_{\nu o}\pm\phi_{\nu \pi}}{\sqrt{2}}
\end{equation}
where $\nu$ represents a spin or a charge index, depending on which particular density
one considers.

\begin{figure}[h]
  \centerline{\includegraphics[width=\figwidth]{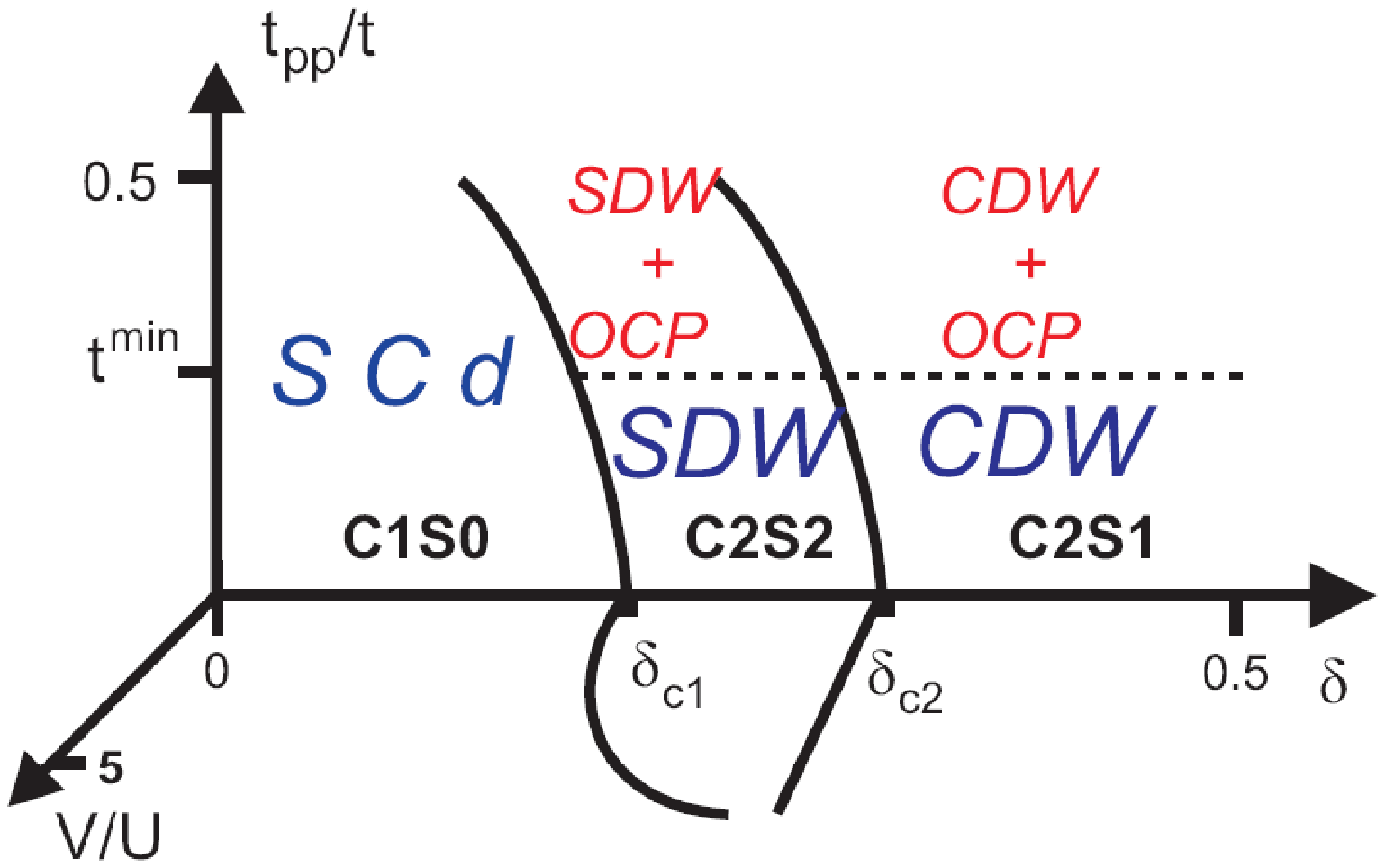}}
  \caption{The phase diagram of two-leg $Cu$-$O$ Hubbard ladders versus doping
  for $U_{Cu}>0$. Zero doping corresponds to
  the half filled case. The umklapp terms which open a gap in
  the charge symmetric mode are not included here}\label{fig:dopin}
\end{figure}

The interaction part, in fermionic language, is given by
\begin{eqnarray}
 \fl   H_{int} =\sum_{j} (\sum_{m\in Cu} U_{Cu}n_{mj\uparrow}n_{mj\downarrow} + \sum_{m\in O}U_{O} n_{mj\uparrow}n_{mj\downarrow}\nonumber \\
 + \sum_{m\in Cu, n\in O}\sum_{\sigma, \sigma'}V_{Cu-O}n_{mj\sigma}n_{nj\sigma'}) \label{hubbInt}
\end{eqnarray}
The bosonized form of Eq.~(\ref{hubbInt}) contains terms of the
density-density type. As we mentioned above, these are included in the $\hat{K}$ matrix,
and hence they are treated exactly in this procedure. Since the form of
$\hat{K}$ depends on the basis in which the densities are
expressed, the Hamiltonian will take the simple form
Eq.~(\ref{eq:Hbozon}) in the eigenbasis of the matrix $\hat{K}$.
The remaining interaction terms yield non-linear cosines and
these are the ones for which the RG procedure is required. In
Refs.~ \cite{chudzinski_ladder_rapid,chudzinski_ladder_long} we
showed that the eigenbases for the spin and for the charge
modes rotate during the RG flow. Two fixed points were found,
namely $B_{+-}$ and $B_{o\pi}$. Interband physics dominates in
the former case (mixing of the $o$ and $\pi$ bands) and
intraband physics, in the latter case. Close to half filling
(low doping regime), $B_{+-}$ is the fixed point basis for the
spin and the charge modes. Both spin modes are gapped, and so
is one of the charges modes. Using the notation of Balents and
Fisher \cite{balents_2ch} the ladder is in the \emph{C1S0}
state (\emph{CnSm} denotes $n$ ($m$) gapless charge (\emph{C})
(spin \emph{S}) modes). For an intermediate range of dopings,
$B_{+-}$ is the fixed point basis for the spin variables and
$B_{o\pi}$ the fixed point basis for the charge variables. This
is the \emph{C2S2} regime when all modes are gapless. The
decoupling of the spin and of the charge eigenbasis is
responsible for this quantum critical state. For higher
dopings, $B_{o\pi}$ is the fixed point basis for the spin and
the charge modes. The ladder is in the \emph{C2S1} phase where
the $o$ spin mode is gapped. The corresponding phase diagram is
shown in Fig.\ref{fig:dopin}.

\subsection{Bosonization of a single impurity}

\begin{figure}
  \centerline{\includegraphics[width=\figwidth]{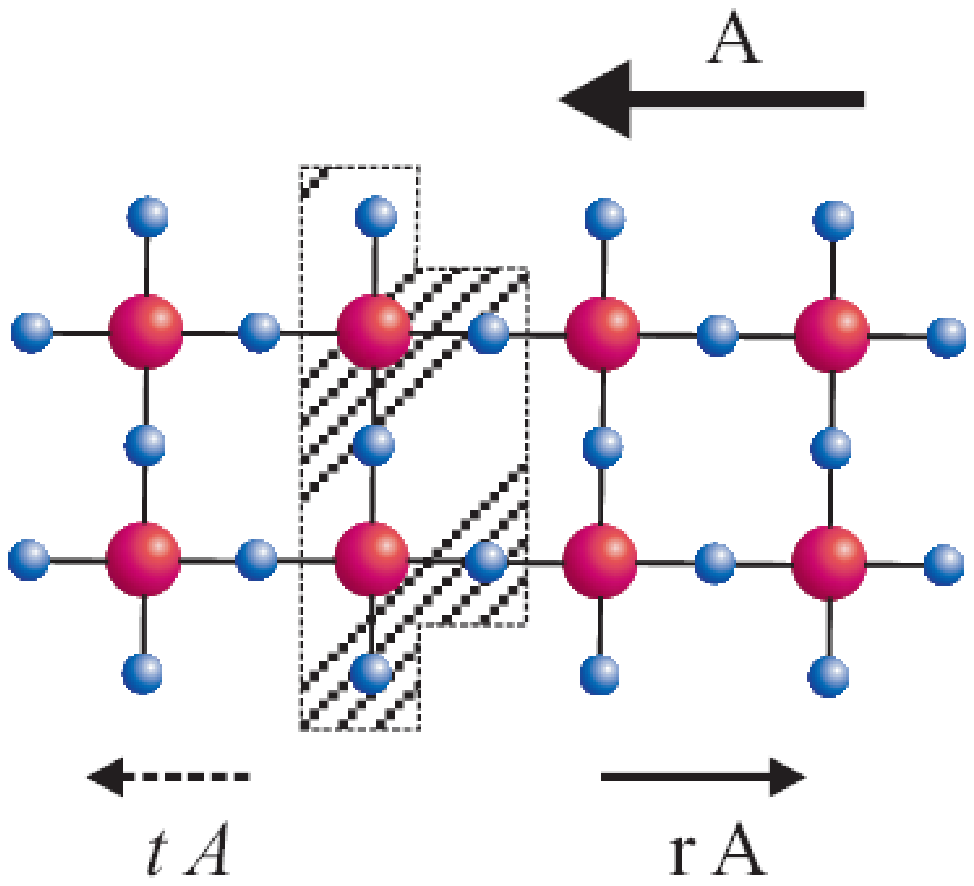}}
  \caption{ A schematic rendition of the scattering event caused by a single impurity: a left moving wave,
  denoted by A,
 impinges upon
  the shadowed region, where an impurity is located. A portion, r A,  is backscattered (the process
  described by a reflection matrix), and the rest, t A, is transmitted to the other side. For moderate or high dopings,
   the latter process only takes place at non-zero temperature.}\label{fig:imp_gener}
\end{figure}

Let us consider a two-leg $Cu$-$O$ ladder with a single defect
located in the unit cell positioned at $x=0$. It could be a
vacancy, an interstitial, or even a distortion caused by the
electrostatic potential of some external source (i.e not located
on the ladder). A sketch of such impurity is shown in
\fref{fig:imp_2types} and its generic scattering effect is
depicted in \fref{fig:imp_gener}.
 Such a defect obviously
affects the motion of carriers along the ladder. In 1D, there
are simply two possibilities for scattering on an impurity:
forward and backward. In  fermionic language this leads to a
contribution of the form
\begin{equation}\label{eq:imp_ferm}
\fl    H_{imp}= \sum_{q\sim
    0}V_{for}(x=0)\sum_{k}\psi_{k+q}^{\dag}\psi_{k}(0)+
    \sum_{q\sim 2k_{F\alpha}}V_{bac}(x=0)\sum_{k}\psi_{k+q}^{\dag}\psi_{k}
\end{equation}
The two terms, in the rhs of Eq. (\ref{eq:imp_ferm}) are totally
decoupled and thus can be treated independently. The strategy is
to rewrite them in a bosonic phase field language, using the
appropriate substitution

\subsubsection{Forward scattering}

In the bosonic representation, we find that forward scattering
acts like an additional chemical potential: $V(0)
(\frac{1}{\pi}\nabla\phi(x=x_{0}))$. It is well known that this
effect can be accounted for by shifting the phase fields in the
spin sector by an amount proportional to the value of the
potential: $\Delta\phi_{i}\simeq\frac{K}{V_{i}}\eta(0)$, where
$\eta(0)$ is the strength of the forward scattering on an impurity
sitting at x=0. We conclude that, for a single impurity problem,
the expection value of the field at x=0 remains unchanged; it does
not influence the analysis of the backscattering part. In
principle, forward scattering effects may thus be omitted.
However, a proviso is in order.

As we are dealing with two bands, We get three independent shifts
(two  from intraband and one from interband forward scattering);
these can be factored out, with the help of the appropriate basis
rotation. Anticipating the analysis of backscattering terms, where
a similar rotation of the basis will be introduced, the following
statement can be made: as long as forward and backward interband
scattering are equal (which is a reasonable assumption) there
won't be any inconsistency in our procedure. This consideration is important,
for example, for the calculation of correlation functions at
finite temperature, where the OBC fixed point is not reached yet.
In the discussion of backscattering, we may distinguish two
limiting behaviors: reflection from a non-magnetic boundary or
coupling to a spin state localized around the impurity. We
devote most of our attention to the former case, since it corresponds to
the LL regime that we are focusing on in this work.  For the spin
(bound) state case, Kondo-like forward scattering terms must be taken into account.

\subsubsection{Backward scattering}

The second part, in the rhs of Eq.~(\ref{eq:imp_ferm}), is more
involved, because, in the bosonic representation, it produces
non-linear terms. In the case of a two-leg ladder, the situation
is even more complicated than for the single chain, since there
are now two different types of backscattering processes - a
reflection within a given band and a reflection from one band
to the other.

The intraband contribution reads
\begin{eqnarray}
 \fl   H_{o}^{(imp)}=V_{o}\cos(\phi_{so})\cos(\phi_{co})|_{x=0}= \nonumber \\
    V_{o}\cos({{\phi_{s-}+\phi_{s+}}\over{\sqrt{2}}})\cos({{\phi_{c-}+\phi_{c+}}\over{\sqrt{2}}})|_{x=0}\label{intrao}
\end{eqnarray}
and
\begin{eqnarray}
 \fl   H_{\pi}^{(imp)}=V_{\pi}\cos(\phi_{s\pi})\cos(\phi_{c\pi})|_{x=0}= \nonumber \\
    V_{\pi}\cos({{\phi_{s-}-\phi_{s+}}\over{\sqrt{2}}})\cos({{\phi_{c-}-\phi_{c+}}\over{\sqrt{2}}})|_{x=0}\label{intrapi}
\end{eqnarray}
The interband contribution, expressed in the $B_{+-}$ basis
(appropriate to treat band-mixing processes) is
\begin{equation}\label{inter}
    H_{o\pi}^{imp}=V_{o\pi}\cos(\phi_{s+})\cos(\phi_{c+})\cos(\theta_{s-})\cos(\theta_{c-})|_{x=0}
\end{equation}

The influence of these non-linear terms will be discussed in the next section,
using an RG analysis. Their relevance signals that backward
scattering is strong enough to ``cut'' the ladder. We will show that
this entails important changes in the expression of correlation
functions.

\section{Field theory for the impurity problem}\label{CFT}

In this section we describe the low energy properties of the
ladder in the presence of an impurity. Two steps are required:
first, one should determine, using RG, the fixed points of the
problem and solve an open boundary problem; second,  from the
fermionic boundary conditions (reflection and mixing amplitudes)
one should determine the proper boundary conditions for the
bosonic phase fields. This is precisely the order in which the
present section is organized. Within each of its subsections we
consider in sequence the \emph{C1S0}, \emph{C2S2} and \emph{C2S1}
phases, which were obtained in the case of a defect-free ladder.

\subsection{RG flows and fixed points}

The RG calculation for the single impurity problem proceeds in two
steps: first one takes the unperturbed ladder and adds the
impurity backscattering potential to check its possible relevance.
When the impurity potential is relevant, one makes an educated guess about
the fixed point of the RG  flow. The
second step consists in checking the stability of this strong
coupling fixed point.

\subsubsection{Weak coupling}

The RG differential equations for the backscattering potential
can be derived using the standard procedure of integrating out
high energy degrees of freedom . One should simply bear in mind
\cite{Kane_RG1,Furusaki_RG2} that, since, $V$ is a local
process at $x=0$, the scaling variable is simply the time
coordinate. The RG flow can be monitored so long as $V$ can be
treated as a perturbation. Two important technical remarks are
in order, before we write down the RG equations and discuss
their solutions. They play an important role in the subsequent
analysis.

\begin{figure}
  \centerline{\includegraphics[width=\figwidth]{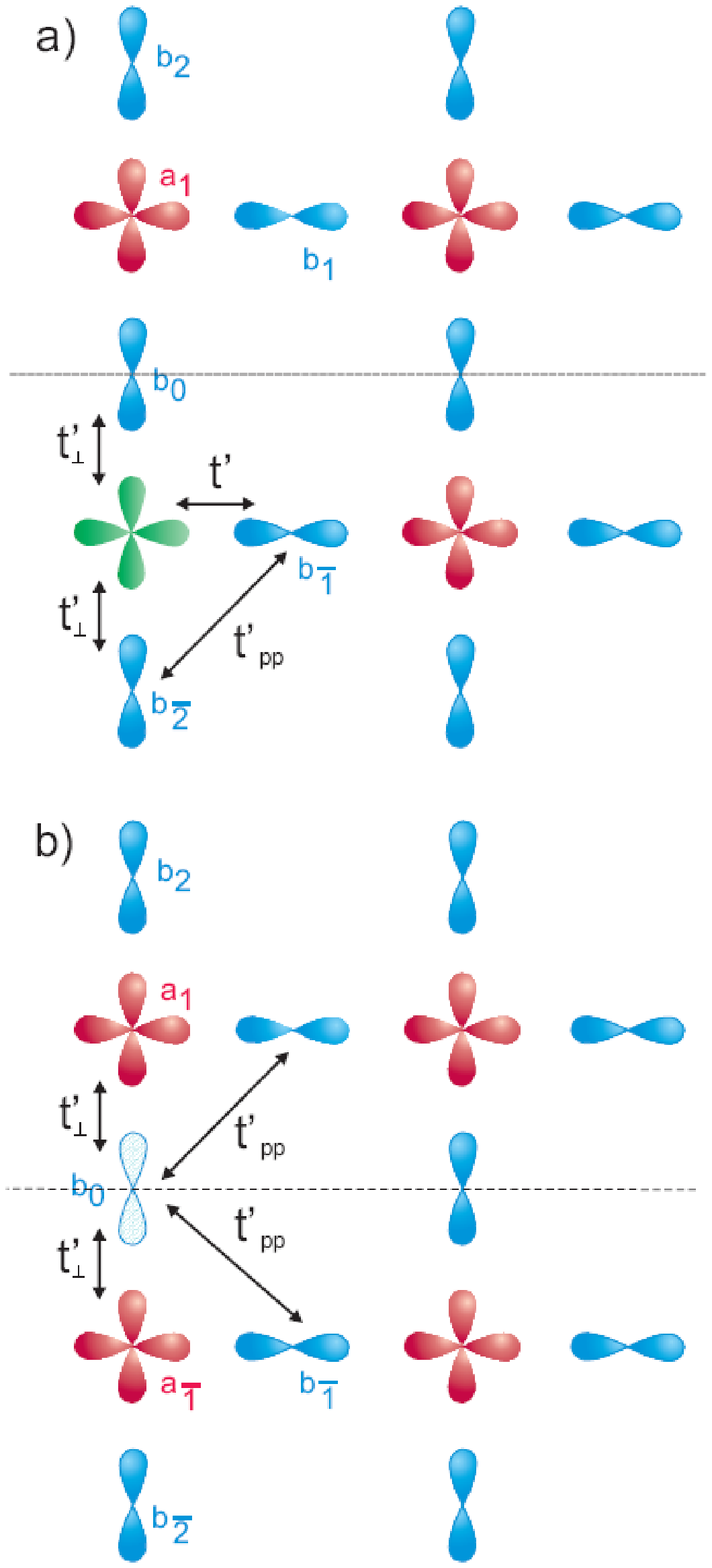}}
  \caption{An atomic view of the unit cell where a defect (in green) is located.
  In an elementary cell, an impurity can be either ``asymmetric'' a) or
  ``symmetric'' b). A ``symmetric'' defect preserves the symmetry with respect to the axis along
  the ladder, shown as a dashed line and denoted $\sigma$ in the text. The hopping amplitudes which
  are affected by the presence of the impurity are shown in each case}\label{fig:imp_2types}
\end{figure}

One concerns the initial conditions that one chooses at the start
of the flow. In the unit cell of the $Cu$-$O$ ladders, there are
several atoms, and hence various possible configurations, when an
impurity is included. One
could consider the substitution of one of the $Cu$ atoms
(the $a_{1}$ d-orbital in
\fref{fig:imp_2types}) by $Zn$ (a non magnetic ion) which creates
a highly asymmetric defect, as shown in
\fref{fig:imp_2types}(a). Alternatively, an oxygen atom
could move away from its equilibrium position, causing a
distortion, or even could be removed from the structure. In that
respect, the on-rung, central (p-orbital indicated as $b_{0}$)
oxygen atom stands out: this type of site is encountered in ladders
and in two-dimensional (2D) cuprate materials. Such a term
preserves the reflection symmetry with respect to the two legs
that one has in the case of a non disordered ladder. This is
shown in Fig.\ref{fig:imp_2types}(b). The outer and on-leg oxygen
sites ($b_{2}$ and $b_{1}$ orbitals respectively) are a bit more
difficult to analyze, because in real systems, where there is
usually some residual coupling between neighboring ladders, they
introduce some mixing between ladders. Distortions that extend
over a few unit cells or removing of apical oxygen can be thought
of as symmetric obstacles. The reflection amplitudes for the different
types of impurities will be evaluated in Sec.~\ref{resultats}.

Symmetry (or lack of thereof) with respect to the exchange of
the two legs of the ladder is a critical issue.  The
corresponding symmetry axis is noted $\sigma$. The asymmetric
case induces a strong initial band mixing process. By contrast,
a defect that sits exactly at the position of an on-rung oxygen
atom does not produce an initial band-mixing. This shows that
the presence of oxygen atoms in the unit cell impacts the
symmetry breaking properties of the impurity. Of course, one is
required to check whether or not this property is preserved
during the flow. If it is not, i.e if band mixing, backward
scattering, terms become relevant during the flow, one may
conclude that symmetry breaking has occurred. Thus the fixed
point ground state phase should be strongly asymmetric. More
generally, the RG procedure may enhance or lower the local
symmetry of the system (by suppressing or reinforcing the
influence of interband backscattering). Such direct link
between RG and symmetries is often postulated, but it is not
always as clearly seen as in the present case. For the $Cu$-$O$
ladder, it shows already in real space: because of symmetry
breaking, the dominant instability which arises close to
impurity should promote a difference between the two legs.

We introduce here such type of instability, as it will recur
frequently in the course of our paper. An obvious example, in the
particle-hole channel,  is the so called $\pi$-density wave
($\pi$-DW) which takes the following forms in language of fermions
(operator $\alpha_{+/-,\mu,\sigma'}$ with chirality, band and spin
indexes) and bosonic fields:
\begin{eqnarray}
\fl    O_{\pi CDW}\sim\sum_{\mu}\sum_{\sigma\sigma'}
\alpha^{\dag}_{-\bar{\mu}\sigma}\delta_{\sigma\sigma'}\alpha_{+\mu\sigma'}= \nonumber \\
\cos\phi_{c+}\sin\theta_{c-}\cos\phi_{s+}\cos\theta_{s-}-
\sin\phi_{c+}\cos\theta_{c-}\sin\phi_{s+}\sin\theta_{s-}\label{eq:zig-zag_def}
\end{eqnarray}
where $\mu =$\{$o$, $\pi$\}, $\bar{\mu}=o$ ($\pi$) if $\mu =\pi$
($o$), and the $-$ and $+$ symbols in the l.h.s of Eq.
(\ref{eq:zig-zag_def}) refer to left and right moving fermions,
respectively. In real space it corresponds to a characteristic
pattern, which will be called  ``zig-zag pattern'' in the course of
the paper. 

The second remark concerns the impact of the impurity potential on
the RG flow of the LL parameters. Since a single impurity may not
change the thermodynamic properties of the system we can use the
fact that the flows of $K_{i}(l)$ and $g_{i}(l)$ are not affected
by the presence of the impurity far enough from it, and one can
simply use results known for the defect-free bulk. Since we are
working in a regime when the initial values of the backscattering
terms are much stronger than that of interactions (but such that
they can still be treated as perturbations), the RG step should be
done as follows, in order to obtain the proper flow for
$V_{i}(l)$:  one should take the functions $K_{i}(l)$ and
$g_{i}(l)$ that were obtained in the absence of disorder and plug
them into the differential equations obeyed by the $V_{i}(l)$.
Inserting instead the fixed point values $K_{i}^{*}$ and
$g_{i}^{*}$ would work in the opposite limit of strong
interactions. This  procedure is meaningful provided we stay at
energy scales above the gaps. At the start of the RG process, the
initial strength of the backscattering potential is assumed to be
small, but one  has to remember that the smaller it is, the longer
it takes to renormalize it to strong values (where one enters the
tunelling regime). In the following RG analysis, we assume that
the initial strength of the impurity is strong enough to drive the
system quickly towards the fixed point, whenever the impurity
potential is a relevant perturbation. This ensures that if the
impurity is relevant, then the effect induced by it will be
visible (below some reasonable temperature/energy scale). Since
the $V_{i}$ renormalize faster than the $K_{i}(l)$ and the
$g_{i}(l)$, it also ensures that there exists an energy window
where the strong coupling (tunelling) analysis of the impurity is
applicable. This requirement is not fulfilled when defects are
located on outer oxygen sites, or when small distortions of the
bond lengths occur, so these situations have to be excluded from
our analysis.

With the above mentioned considerations in mind, we now proceed to
implement the RG procedure. We take into account the rotation of
the eigenbases, and, depending on their fixed point, we write
different sets of equations.  We perform the analysis starting
from the strongly doped ladder limit, and work our way towards the
half filled case.

\paragraph{C2S1 phase:}
at high doping, in the \textbf{C2S1} phase, the intraband basis
($B_{o\pi}$) is relevant and the differential equations
describing the flow are
\begin{eqnarray}
\frac{dV_{o}}{dl} &=& V_{o}(2-K_{so}(l)-K_{co}(l)) \nonumber\\
\frac{dV_{\pi}}{dl} &=& V_{\pi}(2-K_{s\pi}(l)-K_{c\pi}(l)) \nonumber\\
\frac{dV_{o\pi}}{dl} &=& V_{o\pi}(2-\frac{1}{4}(K_{so}^{-1}(l)+K_{s\pi}^{-1}(l)+K_{so}(l)+K_{s\pi}(l) \label{eq:RG_C2S1_o}\\
                   &  &+K_{co}^{-1}(l)+K_{c\pi}^{-1}(l)+K_{co}(l)+K_{c\pi}(l))) \nonumber
\end{eqnarray}
One sees that the $V_{o\pi}$ backscattering term is exactly marginal to
second order in perturbation. For the intraband backscattering:
$d_{0}=(K_{so}(l)+K_{co}(l))<2$, so $V_{0}$ is relevant; $d_{\pi}=(1+K_{c\pi}(l))\leq 2$.
More specifically, it is less than $2$ only at the beginning of the
flow) so $V_{\pi}$ is weakly relevant or marginally relevant.

In the large doping case, the interpretation of the results of
the RG analysis is quite straightforward. The relevant
operators scale to infinity which signals that the intraband
backscattering terms $V_{o}$ and $V_{\pi}$ cut the ladder when
the temperature goes to zero. One may generalize arguments used
for spin chains and assume that the system flows towards the
open boundary fixed point. The only difficulty is how to treat
the $V_{o\pi}$ backscattering. Although it is marginal (i.e not
relevant) one cannot simply neglect its influence, especially
we cannot say that it will scale to zero at the fixed point of
the flow. This bears upon the important issue of whether there
is a single, stable, fixed point or a line of fixed points
(controlled by $V_{o\pi}$), which cannot be excluded by the
weak coupling RG analysis. Nevertheless, contrarily to the case
of the low doping limit, a solution can be found in the
\textbf{C2S1} regime.

\paragraph{C2S2 phase:}
at intermediate dopings, the analysis of the gapless \textbf{C2S2}
phase is also challenging. First of all, the charge sector is
described by the intraband basis ($B_{o\pi}$) while the
spin sector, as for lower dopings, is described by the total
transverse one ($B_{+-}$). Besides, because of spin rotational
invariance
all the LL parameters $K_{s}$ are equal to one. This simplifies the RG
equations to
\begin{eqnarray}
\frac{dV_{o}}{dl} &=& V_{o}(1-K_{co}(l)) \nonumber \\
\frac{dV_{\pi}}{dl} &=& V_{\pi}(1-K_{c\pi}(l)) \label{eq:RG_C2S2}\\
\frac{dV_{o\pi}}{dl} &=& V_{o\pi}(1-\frac{1}{2}(K_{co}^{-1}(l)+K_{c\pi}^{-1}(l)+K_{co}(l)+K_{c\pi}(l))) \nonumber
\end{eqnarray}
For these equations the $V_{o\pi}$ backscattering term is exactly marginal  to
second order in perturbation. For the intraband backscattering terms:
$d_{0}=K_{co}(l)<1$, so $V_{0}$ is relevant and $d_{\pi}=K_{c\pi}(l)<\simeq 1$ so $V_{\pi}$ is weakly relevant.

However one should be careful when interpreting this result.
The fact that the spin sector, which is crucial both at low
doping (spin screening) and at high doping (spin gap), does not
appear to influence the relevance of backscattering is quite
puzzling. It is known \cite{Le_Hur_imp, Granath_mulKondo,
Eggert_RGcomp} that for a critical SU(2) invariant phase, the
band mixing term $V_{o\pi}$ may cause important (in fact
dominant) logarithmic corrections. These corrections favor an
instability (spin screening) of the same nature as that which
pertains to the \textbf{C1S0} phase. In this respect the
physics is similar to that discussed before for the low doping
regime. Analytical studies of this problem were devoted to the
case of spin ladders, when charge modes are frozen out. Here
charge modes are present and the fact that their fluctuations
are described in the $B_{o\pi}$ basis could potentially
suppress the influence of the $V_{o\pi}$ backscattering term.
Thus the discussion of this intermediate phase will be
postponed until the end of this paper, because we are not in a
position to make exact predictions about the strong coupling
fixed point in this regime, at that stage.

\paragraph{C1S0 phase:}
For the \textbf{C1S0} phase (low doping case), the total/transverse basis is relevant
and the differential equations describing the flow are given by
\begin{eqnarray}\label{eq:RG_C1S0_1}
\frac{dV_{o/\pi}}{dl} &=& V_{o/\pi}(2-\frac{1}{2}(K_{s-}(l)+K_{s+}(l)+K_{c-}(l)+K_{c+}(l))) \\
\frac{dV_{o\pi}}{dl} &=& V_{o\pi}(2-\frac{1}{2}(K_{s-}^{-1}(l)+K_{s+}(l)+K_{c-}^{-1}(l)+K_{c+}(l)))\nonumber
\end{eqnarray}
The $V_{o\pi}$ backscattering term (band mixing caused by an
impurity) is the simplest to analyze. $K_{c+}$ is always much
smaller than one, $K_{s+}$ may decrease but not increase and
$K_{c-}$ increases but may not decrease, so this backscattering
operator, if it is present, is the most relevant one. The analysis
pertaining to the intraband backscattering terms $V_{o}$ and
$V_{\pi}$ is a little bit more involved, as
$d_{0}=d_{\pi}=\frac{1}{2}(K_{1}+K_{2}+K_{3}+K_{4})$ is initially
smaller than $2$ but that, in the course of the flow, it exceeds
this value. This suggests a non monotonous evolution of the
potential during the RG procedure: first it increases and then it
decreases (so the overall amplitude of its variation will depend
on the scale which chararacterizes the approach of $V_{0\pi}$
towards unity).

Thus, the band mixing process, which takes us away from the simple four
LL mode generalization of the spinful chain problem, dominates the
low energy physics. In addition, intraband RG gives a
non-monotonic behavior, so the initial conditions play a
critical role in this case. The only statement that can be safely made is
that the system flows away from the unperturbed situation. In
other words,  perfect transmission is an unstable fixed point of
the flow in the low doping regime. We need to assess the regime
towards which $V_{i}$ flows. Within the range of validity of our
treatment, we cannot produce a rigorous solution, but we provide below
some reasonable indications.

Some statements can also be made as regards the fixed point
symmetry properties. If we assume that the initial conditions of
the flow are symmetric, then because $V_{o\pi}$ is strongly
relevant, any deviation from this ideal position would cause
band-mixing processes. One can see that an exactly symmetric defect
is also an unstable fixed point in the phase space of
intra- and inter- band scatterings. The system flows towards a strongly
asymmetric situation, where fluctuations affect each leg
differently.

These findings have an impact on a symmetry analysis of the
impurity problem: for high dopings we do not predict any
enhancement of band-mixing at the fixed point; then a zig-zag
pattern will be much more visible for $Cu$ atom substitutions than
for  other, more symmetric, types of distortion. For all cases,
intraband physics dominates, and one may apply the BCFT procedure.

To summarize, the RG analysis is not able to
give a full answer to the backscattering relevance question, but
we are still able to distinguish two regimes: at low dopings,
interband scattering dominates, whereas, at high dopings, intraband
reflection prevails. In order to apply BCFT (for the bosonic
fields) one needs to have stronger intraband backscattering than
band-mixing, at the fixed point. Thus, strictly speaking, our
method is well suited to the high doping regime. Before setting to
apply BCFT one should check the stability of the suggested fixed
points.

\subsubsection{Strong coupling}

Performing an analysis in the strong coupling regime requires
some knowledge of the characteristics of the low temperature
fixed point. For single chains this is an easy task, since a 1D system is
either cut or not, but as we show below, this is not as straightforward
for ladders. What is more, the nature the fixed point around
which the analysis is performed changes from phase to phase.
Thus, analyzing the effect of the impurity for each phase will
be even more involved than in the weak coupling (perturbative) limit. At the
end of the section we provide some arguments supporting the
possible existence of ``exotic'' fixed points (multi-particle
tunelling processes).

\paragraph{C2S1 phase:}
Let us discuss first the large doping limit, because in this
case the strong impurity fixed point is much easier to
identify. On the basis of the above arguments, one may conclude
that, for the \textbf{C2S1} phase, there is no intermediate
fixed point and the impurity simply scales to the tunelling
limit. We have then established the limits of validity of our
approach, based on CFT. We will come back to this point in the
discussion part (Sec.~\ref{discussion}). We now turn to the
physics at the fixed point. In the case when intraband
backscattering dominates, one can use the duality procedure
 \cite{Kane_RG1} and analyze the flow starting from the strong
coupling fixed point of a weak tunelling between two
semi-infinite ladders. For single chains, it was shown, that
the tunelling problem could be solved using a saddle point
approximation. The sequential tunelling events are described in
terms of a dilute instanton gas. In this dual mapping, the
quadratic part which was previously the free part, is
responsible for interactions between instantons. One obtains
the same partition function, through the simple mapping
\cite{Kane_RG1} $\phi_{i}\rightarrow \theta_{i}$ . It was later
shown \cite{Furusaki_RG2} that this can be generalized to the
case of two diagonal (spin and charge) LL modes. In our case,
when all $K_{i}$ are close to one, the weak/strong impurity
regimes are exactly dual and the boundaries (in $K_{i}$ phase
space) of these two limits coincide. We will come back to this
point at the end of this section. In a similar manner one may
straightforwardly generalize the analysis to the case of four
LL modes. One point has, however, to be taken into
consideration: for the 1D chain, spin and charge are
independent degrees of freedom and, to first order, they do not
mix on the boundary.
By contrast, from the geometry of the ladder, one sees that the impurity usually causes some mixing of the $o$
and $\pi$ bands:
 if the impurity is symmetric with respect to the
$\sigma$ axis, we cannot get any mixing, but in the generic
case, when only one of the chains is cut, the chain, not the
band, basis is appropriate to describe the propagation of
carriers. The chain basis implies band mixing.

In the case of a symmetric barrier, the band description is still
valid and we may define instantons using a saddle point
approximation within each band. It is the intraband instability
which corresponds to this kind of impurity, and the dominant
tunelling process is
described by a term
 which
can be treated to second order in  perturbation (the first
order contribution vanishes)
\begin{equation}\label{tunelRG}
    \int dt_{1} dt_{2}
    \langle \psi_{1o}^{\dag}(x=0,t_{1})\psi_{2o}(x=0,t_{1})
    \psi_{2o}^{\dag}(x=0,t_{2})\psi_{1o}(x=0,t_{2}) \rangle
\end{equation}
This leads to the limiting situation of two independent,
semi-infinite, ladders \emph{1} and \emph{2}. The result of
Ref.~\cite{Furusaki_RG2} can easily be generalized in this
case, and we find an OBC stable fixed point. Of course, even at
high dopings, one cannot \emph{a priori} rule out the
possibility of band mixing backscattering, but we may predict
that the size of the effect in the tunelling regime is similar
to that calculated initially at the beginning of RG procedure.

The case when only one of the legs is cut is a little more
involved. Using the same argument as before, we find that a type
of instability which minimizes the energy (in the sense that it
produces a zero expectation value at the position of the impurity)
is now the $\pi$-charge density wave ($\pi$-CDW). It the only type
of density wave (DW) ordering which mixes the $o$ and $\pi$ bands
and it is defined within each chain, as expected. Then, the
dominant type of tunelling is expected to be of the following
form: $\psi_{1o}(x=0)\psi_{2\pi}(x=0)\sim
O_{\pi-CDW}(t_{1},t_{2})$. One can now make the same exact
calculation (to second order in perturbation) for this kind of
process, or use the dilute instanton approximation for the
$\pi$-CDW instantons to find an equation similar to that for
$V_{o\pi}$, but with the $K_{i}\rightarrow K_{i}^{-1}$ mapping.
Note that here are two approaches to the
tunelling problem: one treating it as a second order perturbation
(see Eq. (\ref{tunelRG})) with known OBC density of states, another using
quasi-classical instanton solutions. Both give the same answer.
One sees then that, for high dopings, this kind of tunelling is
marginal -- it will not destabilize the OBC fixed point -- but it
has to be taken into account in the analysis. It might seem
surprising that, even if the perturbation only affects one of the
legs, we end up at the reflecting fixed point anyway, but an
analogy with the Fano resonance may be useful here
 \cite{Fano_resonance,Bulka_Fanores}. In the Fano problem, we have
two channels: one with weak and one with strong transmission. If
the coupling between the two is strong (as is the case here), the
so- called asymmetry parameter $q$ is equal to zero
 \cite{Fano_resonance} and one observes a dip at the Fermi energy,
which is increasingly pronounced as the temperature is reduced.
Due to quantum interferences between the channels, the conductance
tends to zero with decreasing energy scale (temperature, gate
voltage).

Within the RG framework, we have shown so far that, for the
impurity problem,  there exists a stable line, toward which the
flow takes us, corresponding to an intra- or an inter- band
reflection. It might be desirable to address the issue of the
nature of flow on this line (for instance, if we have a line of
a fixed points), but this lies beyond the scope of the present
work. A remark is in order here. There is always an intrinsic
infrared cut-off, due to some characteristic size (inverse of
the impurity concentration) or to a finite temperature; also,
depending on where the impurity sits inside the unit cell, one
could get marginally relevant scattering processes which
strongly affect the flow. Taking these effects into account, it
is safest to consider more general boundary conditions,
including reflection processes of both types. Hence, we do not
assume that, upon reaching the \emph{true} zero temperature
fixed point, one of these two types of backscattering
disappears.

\paragraph{C2S2 phase}

At this point, we are not yet in a position to discuss the
influence of the impurity for this phase. First, we need to
investigate the low doping regime more closely. But an
important statement is in order; until now we had considered
that, for each doping, the phase in contact with the impurity
was that of the defect-free ladder at that particular value of the
doping; this might not be the case for a critical, gapless,
phase which is usually very sensitive to any perturbation. We
will come back to this important issue in Sec.~\ref{discussion}
(in particular when we discuss the gapless phase).

\paragraph{C1S0 phase:}
Performing exactly the same RG analysis of all possible tunelling
terms as that described above -- in the low doping phase -- shows
that no suppression of intraband instanton events occurs, down to
the lowest temperature. In the charge sector, for temperatures
$T<\Delta_{c-}$, we find $K_{c-}\gg 1$ (while all other $K_{i}\sim
1$), so that the intraband transmission coefficients $t_{o,\pi}$
become relevant, and this makes the OBC fixed point unstable.
Only the interband $t_{o\pi}$ is strongly suppressed which means
that a $\pi-DW$ resonance is produced around the impurity. In
this case, it is desirable to identify the nature of the new,
intermediate, fixed point. Note that the above statement holds only
so long as the $c-$ mode remains gapped (the condition $K_{c-}\gg
1$ is directly connected with this), so it should not be valid in
the massless \emph{C2S2} phase.

The relevance of band-mixing terms for the low doping phase
suggests that the physics is dominated by on-chain boundary
states. The \emph{C1S0} phase has an SCd character, with
singlet on-rung Cooper pairs. Removing one of sites leaves an
unpaired free spin on the other chain. In addition, at half
filling, rewriting the gapped spin sector in terms of Majorana
fermions allows one to identify the boundary state as a triplet
(the condition for the triplet excitation mass, $m_{t}>0$, must
be fulfilled \cite{Lecheminant_Majorana}). Finally, from the
dispersion relation of a single $Cu$-$O$ chain (in a nearest
neighbor tight binding model approximation), one concludes that
the on-chain resonance can only exist in a low doping energy
range. Extending this arguments, which holds at half-filling
(strong interactions regime), to the entire C1S0 phase,
suggests that for weak doping the fixed point is different from
that describing a simple reflection on the boundary. Instead a
bound state (exponentially decaying into the bulk) with a spin
$1/2$ forms around the defect and the system flows to a
different fixed point. This behavior is not quantitatively
captured, in the framework of our approach (for other examples
of such situation, see Ref.~\cite{Eggert_RGcomp}), so we will
only make
some qualitative predictions. 

To that end, it is worthwhile performing an RG
analysis of the various Kondo-type couplings with the boundary
spin-state.
The general scattering operator on the impurity
spin $\vec{S}$ has the following form:
$$
H_{b-imp}=\sum_{\alpha\alpha'}J_{K}^{\alpha\alpha'}(\psi_{\alpha}^{\dag}\vec{\sigma}\psi_{\alpha'})\vec{S}
$$
where $\alpha$ is the band index (we can have either
intra- or inter- band scattering in our system). The band part
$(\psi_{\alpha}^{\dag}\vec{\sigma}\psi_{\alpha'})$ can be
rewritten in bosonic language, giving rise to terms proportional to
$J_{K}^{\alpha\alpha'}\exp(\imath\phi_{Rs+/-}(x=0))S_{i}$. Here,
$i=z, +, -$, depending on whether the $z$ component or spin-flip processes of the impurity spin are 
concerned. Following Affleck, the impurity spin only couples to
one chiral (right going in our case) spin field,
since backscattering on the boundary is relevant.

Some details regarding the treatment of the above Kondo terms
are presented in \ref{Kondo}. The most important difference
between inter- and intra- band terms is the presence of a
$\cos(\theta_{Rc-}(x=0))$ term in the former case.
An RG analysis of all possible $J_{K}^{\alpha\alpha'}$ can be
performed assuming OBC (relevance of $V_{o\pi}$). Because of the
presence of a charge transverse term ($K_{c-}>1$), $J_{K}^{o\pi}$
is strongly relevant while the other couplings are marginally
relevant. As the temperature is lowered further, we approach the
energy scale for which $K_{s}<1$ (which corresponds to the opening
of a spin gap). In this regime all the $J_{K}^{\alpha\alpha'}$
couplings become irrelevant. Hence, we may write an effective
Hamiltonian for the spin degrees of freedom below the charge gap
$\Delta_{c-}$, but above the spin gap energies $\Delta_{s}$.
 It describes interband Kondo scattering:
\begin{equation}\label{eq:Kondo_eff}
 \fl   H_{b-imp}=J_{K}^{o\pi}\cos(\phi_{Rs-}(x=0))S^{z}
    +J_{K}^{o\pi}(S^{+}\exp(\imath\phi_{Rs+}(x=0))+h.c.)
\end{equation}
Let us point out that, in contradiction with the usual case, it
is the total spin field $\phi_{Rs+}$ which is responsible for
the impurity spin flip processes. The form of this spin-flip
term suggests that (as for a single-channel Kondo problem) bulk
carriers are able to screen the impurity spin at zero
temperature. In addition, the term coupled to $S^{z}$ is not of
the density-density type, which is why it has a non-linear
cosine form rather than the customary gradient form.
Hamiltonian (\ref{eq:Kondo_eff}) can be mapped onto a resonant
level model, albeit with an extra interaction term stemming
from the $z$ component of the spin.
The above RG arguments suggest the existence of an additional,
stable, intermediate fixed point in the low doping regime. It
is easy to show that the $J_{K}$ couplings are relevant only if
$B_{+-}$ is the fixed point eigenbasis, which is consistent
 with the existence of (impurity) spin states only at low dopings. We will
explore the nature of this new state in more detail, in
Sec.~\ref{discussion}.

\paragraph{Exotic tunelling processes}

To complete our analysis of the strong coupling limit, we need
to check if additional tunelling processes could be generated
in higher order. For the case of the spin chain,
Ref.~\cite{Furusaki_RG2} included the possibility of double
spinon or double holon tunelling events. Such processes
dominate if the relevant LL parameter satisfy $K_{i}<1/2$. In
our model, there are four modes and so, obviously, many more
processes of this type. Using a path integral Monte Carlo
technique, the boundaries between such tunelling regimes, for
all dopings, were investigated \cite{hamamoto_piMC} and the
crossover regime between the weak and the strong impurity cases
were determined. No additional intermediate fixed points was
found. We would then conclude that in our model, as long as all
$K_{i}\sim 1$, this kind of exotic tunelling process cannot
make the fixed point unstable. One might of course worry that
$K_{i}\sim 1$ is a property that is connected with the weak
coupling treatment of interactions. Fortunately, at high
dopings, we can waive this concern. In that limit, the $K$
parameters in the $\pi$ band have an intrinsic tendency to
renormalize towards one, which means that no spinon-holon
separation can be expected around the impurity. Regarding the
$o$ band the situation is more complicated: one finds that
$K_{so}$ renormalizes towards the Luther-Emery point
($K_{so}=1/2$), and, at the same time, that interactions open
up a gap in this mode (which corresponds to the freezing of
spinons). On the other hand interactions push $K_{co}$ strongly
below one; it is then quite difficult to reach a regime where
double holon tunelling will make the fixed point unstable. From
the above mentioned numerical analysis \cite{hamamoto_piMC} it
appears that the region of the phase diagram where these exotic
tunelling events dominate is even more suppressed, and, for a
given value of $K_{so}=1/2$, the condition $K_{co}>2$ would
need to be fulfilled.

The situation is not as clear in the low doping regime,
when the $K_{i}$ of the various modes flow either to zero
($i=s-, s+$) or to infinity (i=c-). In principle it is then
possible to have $K_{c-}\gg K_{s+/-}$, so by analogy with
the $\cos(2\phi_{s})$ term introduced by Nagaosa, we may construct interband processes of
the type $\cos 2(\theta_{s+}+\theta_{s-})$. The condition for these double spinon tunnellings in the interband channel to occur
may be fulfilled. However one has to remember that while
$K_{s+/-}\rightarrow 0$ these two spin modes acquire gaps
simultaneously. Thus the perturbative RG description in the LL
framework cannot be valid anymore. The only firm conclusion that we
may reach regards the stability of the perfectly reflecting fixed
point: even if we are not in the single instanton tunnelling regime,
double spinon tunnelling processes may
give rise to the other instability.
According to the above considerations, double holon or double
spinon tunelling events can dominate if their LL parameters are
very different. This can indeed occur, since spin-charge
separation is strongly enhanced by the interactions. One might
thus interpret the regions of Furusaki-Nagaosa exotic
tunelling, as ones where interactions act very differently on
the spin and the charge sectors (giving rise to very different
K parameters). This scenario could prove even more relevant
when one turns to the quantum critical \emph{C2S2} phase.
Spin-charge separation holds all the more since the eigenbases
characterizing the spin and charge modes are different.
Fluctuations in these two sectors are fundamentally decoupled.
Higher order tunelling processes could conceivably play a
crucial role in this regime.

\subsection{Boundary conditions for the phase fields}\label{sec:bound}

As was shown in the previous section, there exists a regime for
the two-leg ladder, where the flow of the (single) impurity
potential leads to a perfect reflection of the carriers in the
system. This is the case of the OBC fixed point; at zero
temperature, only half of the ladder is available for the
motion of carriers and correlation functions around the
impurity have to be modified in order to accommodate this
constrain. In the framework of BCFT, this problem is solvable
using a bosonic field language. Thus, for the massless LL
phases, which are the main focus of this paper, correlation
functions and susceptibilities can be obtained in closed
analytical form (in terms of bosonic fields). For a single
chain, when, because of the OBC, only the $x>0$ half-space is
available, the procedure is as follows:  one writes fermionic
operator in terms of the non-interacting chiral bosonic fields
and then one replaces the impurity by a static, scale invariant
boundary condition. This condition, corresponding to charge
conservation on the exactly reflecting boundary, maps
\cite{Eggert_gen1} a right-going field at $x>0$ onto a left
going field at the ``virtual'' position $x<0$
$\phi_{R}(x)=\phi_{L}(-x)$. The theory is then written using
the fields of one chirality ($R$ or $L$) and, introducing the
usual $\phi,~\theta$ fields one gets the required correlation
functions.

This scenario is quite simple for the chain, since there is only one
type of reflection for fermions and hence the condition for
fermionic and bosonic fields coincide. When two bands are involved
the problem is more involved. Indeed, fermions can undergo  intra-
and/or inter- band scattering. In fermionic language, and for
carriers in the $o$ band, one may write
\begin{equation}\label{eq:bound fer}
    \psi_{Ro}=r_{oo}\psi_{Lo}+r_{o\pi}\psi_{L\pi}
\end{equation}
the bosonic fields appear in the exponent of fermionic
operators, so we cannot take the logarithm of both sides, as is
done in the case of a chain. In order to take the band mixing
term into account, we need to introduce an additional step in
the procedure. From the discussion of the previous section it
appears that this troublesome term (in the region of interest)
close to fixed point should be much smaller than the intraband
backscattering.

The aim is now to rewrite this boundary condition in bosonic field language.
First we simply substitute fermionic operators:
\begin{equation}\label{eq:bound bos1}
\fl    \exp(\imath
k_{Fo}x)\exp(\imath\phi_{Ro})=r_{oo}\exp(-\imath
    k_{Fo}x)\exp(\imath\phi_{Lo})
    +r_{o\pi}\exp(-\imath k_{F\pi}x)\exp(\imath\phi_{L\pi})
\end{equation}
Obviously one cannot write boundary condition in the form
$\phi_{R}(x)=\phi_{L}(-x)$ for the fields $\phi_{o/\pi,L}(-x)$.
The idea is then to write the band phase fields as a linear
combination of some other fields $\phi_{o/\pi}=a\phi_{a/b}\pm
b\phi_{b/a}$:
\begin{equation}\label{eq:comb bos}
\fl \exp(\imath(a\phi_{Ra}+b\phi_{Rb}))=r_{oo}\exp(\imath(a\phi_{La}+b\phi_{Lb}))
+r_{o\pi}\exp(\imath(a\phi_{Lb}-b\phi_{La}+\delta))
\end{equation}
and to find  coefficients which will allow us to factorize the
right hand side of (\ref{eq:comb bos}). The final form of the
right hand side will then be
\begin{equation}
r_{a}\exp(\imath(a\phi_{La}))r_{b}\exp(\imath(b\phi_{Lb}))
\end{equation}
We have some flexibility in choosing both the $r_{a,b}$ and $a,b$
coefficients (only their products are fixed), so using some
algebra, we can find a solution to the problem namely the functional dependence
of  $a$ on $r_{o}$, $r_{\pi}$, $r_{o\pi}$. The factors $r_{a,b}$ can
be absorbed using additional phase shifts for each field. To
lowest order approximation (taking $r_{o\pi}$ much smaller than
the other $r_{i}$), we find that $b\sim r_{o\pi}/\bar{r}$ where $\bar{r}=(r_{o}+r_{\pi})/2$ (the avarage intraband backscattering).

\begin{figure}
  \centerline{\includegraphics[width=\figwidth]{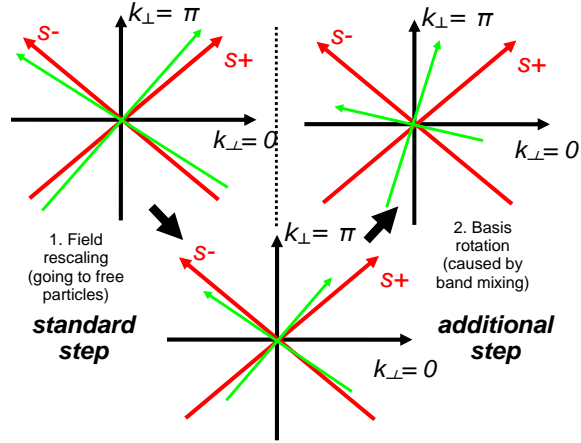}}
  \caption{The two steps necessary to derive bosonic fields
  boundary conditions: the standard rescaling of the fields to
  absorb interactions and the additional basis rotation to
  take into account band mixing. The boundary condition is
  written in terms of these final fields.\label{fig:2step}}
\end{figure}

Assuming that the fields $\phi_{a,b}$ are independent, we may
write separate boundary condition for each of them. Taking the
logarithm of Eq. (\ref{eq:comb bos}), we find the required
boundary conditions for the bosonic phase fields:
\begin{equation}\label{bound bos2}
    \phi_{La}(x)=\phi_{Ra}(-x)+\delta_{a}
\end{equation}
Note that these conditions are written in terms of
fields defined in the new basis.
Band mixing causes a rotation, and this additional step has been
taken into account in Eq. (\ref{bound bos2}). It was doable
since we are working with the free fields. Correlation functions,
in that situation, are the same for all bases, hence we choose the
one that includes band-mixing.

One may wonder what would happen if one used this result for
the Green functions at the LL boundary and then if we plugged
them back into the equations describing the tunelling regime.
For a single mode LL, it is known \cite{giamarchi_book_1d}
that, using the boundary density of states to compute the
energy dependence of the tunelling rate, yields the same result
(a $K^{-1}$ exponent in the power laws) as was found in the
above described instanton analysis. These exponents are taken
into account in the derivation of boundary conditions by
bosonic fields rescaling (the first step in
Fig.\ref{fig:2step}). A similar procedure can be used to
confirm the second step introduced by us (bosonic modes basis
rotation). The last property can be seen to hold when one
evaluates tunnelling rates in the general case of arbitrary,
diagonal density of states \cite{Krive_inter} (for example
rotated by the presence of boundary band mixing). This result
-- linear dependence on band mixing -- is in full agreement
with our RG analysis which shows that band asymmetry suppresses
interband tunelling; so we do not need to perform further
cross-checks, here.

We summarize here the main result of this subsection: the
band-mixing process was included in the bosonic boundary
conditions, showing that, by and large, the single mode LL
solution is stable with respect to this type of perturbation.

\section{Results}\label{resultats}

Let us now present the results of our calculations. In some
regimes the behavior will be similar to that observed in the
case of a single chain \cite{kakashvili_LSW, Eggert2}, but for
the case of the ladder, new phenomena also occur. Furthermore,
we underscore the connection between the diagonal basis found
for disorder-free ladders and the nature of the regime which
appears in the vicinity of a non-magnetic impurity.

Once boundary conditions are known, we may evaluate all correlation
functions of fermionic operators. Thus some experimentally
testable predictions of our BCFT results can be made. We will
determine quantities which are relevant to STM and
to NMR spectroscopies.  STM will provide a local information about
the charge density at some given position in the sample. From NMR
experiments, one can extract some information regarding local
magnetic properties around each atom. So these two techniques have
the ability to give us clues about the states in the vicinity of
impurities. Results given in this section mostly pertain to the
intermediate and highly doped -- free spinon -- regimes. The low
doping regime is beyond the scope of this paper. More
specifically, the results presented below are valid as long as OBC is
a zero temperature fixed point of the RG flow (even if this
behavior is masked by the presence of a gap, as is the case  for
the $o$ band in the \emph{C2S1} phase).  The finite temperature
conformal mapping of the LL is justified as long as the zero
temperature fixed point is well controlled.

\subsection{Green functions}\label{sec:Green}

Let us start with an evaluation of local (in space) Green
functions, which are important for the calculation of local
density profiles. These Green functions can be readily obtained
by rewriting the fermion creation operator in terms of chiral,
free bosonic fields
$\tilde{\phi}_{a/b}^{L}(x),~\tilde{\phi}_{a/b}^{R}(x)$:
\begin{equation}\label{eq:finalfer}
    \psi_{Ro/\pi}(x,t)=\exp(a(F(x,t;K_{o/\pi})))\exp(\pm b(F(x,t;K_{\pi /o})))
\end{equation}
where x is the distance from impurity and the function
$F(x,t;K_{o/\pi})$ is a standard sum of phase fields
parameterized by the proper LL parameters $K_{i}$, similar to
the one derived for the OBC problem in Ref.~\cite{Eggert_gen1}.
These sums can be obtained in two steps from the standard
representation of $\psi_{R,L}$ in terms of bosonic fields
$\phi_{i},~\theta_{i}$: the first step (see
Fig.~\ref{fig:2step}) is the standard field rescaling to
exclude interactions; in the second step one needs to go to the
\emph{a-b} basis in which band mixing is incorporated by the
proper rotation. Then we implement the standard procedure: we
use the boundary condition to introduce the
$\tilde{\phi}_{a/b}^{L}(x),~\tilde{\phi}_{a/b}^{L}(-x)$ fields,
and, at the end, we revert to canonically conjugate fields
$\tilde{\phi}_{a/b}(\pm x),~\tilde{\theta}_{a/b}(\pm x)$ where
correlation functions are known. The procedure is standard
 \cite{kakashvili_LSW, Eggert_stm}, it was even used to fit
numerical results for different types of 1D systems
 \cite{Rommer_BCFT+TMRG}, and the only difference with previous
calculations is the presence of a,b coefficients. Using
Eq. (\ref{eq:finalfer}), allows us to compute all correlation functions
using the bosonization procedure. Spin-charge separation
remains valid, so that correlations are products of charge and
spin parts (for a LL, each piece is a power law with an exponent
determined by $K_{i}$). They are products of a and b parts
(instead of $o$ and $\pi$). The fields $\phi_{i,a,b}$ are similar
to those introduced in the case of non-interacting fermions, since
all interactions were absorbed through $K_{i}$ (or
$K_{i}^{-1}$) rescaling, at the beginning of the procedure. Reverting to
unscaled fields yields additional coefficients $a^{2}$, $b^{2}$ or
$ab$. Note that the condition $a^{2}+b^{2}=1$ has to be preserved,
so that intraband correlators remain unaffected by the additional
basis rotation. The new effect brought by the reflection on the
impurity is the possibility of left/right correlations of movers
in different bands. This will produce some asymmetry between the
two legs of the ladder. This effect causes $a,b$  to play a role,
giving an additional coefficient which changes the amplitude of
the interband correlation functions by an amount $\exp(ab)\sim
r_{o\pi}$.

For one LL mode, using CFT at finite temperature $T
(=\beta^{-1}$), we have
\begin{eqnarray}
\fl F_{\nu}(x,t;V_{\alpha},K_{\alpha})=(\frac{\pi}{\beta
V_{\nu\alpha}})^{(K_{\nu\alpha}+1/K_{\nu\alpha})/4}
|\sinh(\frac{\pi}{\beta}(t+2x/V_{\nu\alpha}))|^{-(K_{\nu\alpha}+1/K_{\nu\alpha})/8-1/4} \nonumber\\
|\sinh(\frac{\pi}{\beta}(t-2x/V_{\nu\alpha}))|^{-(K_{\nu\alpha}+1/K_{\nu\alpha})/8+1/4}
\left|\frac{\sinh(\frac{t\pi}{\beta})}{\sinh(\frac{2x\pi}{V_{\nu\alpha}\beta})}\right|^{-(K_{\nu\alpha}+1/K_{\nu\alpha})/4}\label{corr Egg}
\end{eqnarray}
where $x$ represents the distance from the boundary, and where
we use units such that $\hbar=k=1$. The propagators needed in
the evaluation of correlation functions for the \emph{C2S1}
phase are straightforward generalizations of those pertaining
to the single chain case, and hence we get
\begin{eqnarray}
\fl    \langle\psi_{Lo}^{\dag}(x,t)\psi_{Ro}(x,0)\rangle &=&
\exp\imath
    (2k_{Fo}x+\delta_{o})\prod_{\nu=c,s}F_{\nu}(x,t;V_{Fo},K_{o}) \nonumber \\
    &  &+\exp-\imath (2k_{Fo}x+\delta_{o})\prod_{\nu=c,s}F_{\nu}(-x,t;V_{Fo},K_{o}) \label{corr o}\\
\fl    \langle\psi_{L\pi}^{\dag}(x,t)\psi_{R\pi}(x,0)\rangle &=&
\exp\imath
    (2k_{F\pi}x+\delta_{\pi})\prod_{\nu=c,s}F_{\nu}(x,t;V_{F\pi},K_{\pi}) \nonumber \\
    & & +\exp-\imath (2k_{F\pi}x+\delta_{\pi})\prod_{\nu=c,s}F_{\nu}(-x,t;V_{F\pi},K_{\pi}) \label{corr}
\end{eqnarray}
These expressions allow us to reach very similar conclusions to
those found previously for the single chain \cite{Eggert_stm}. One
easily recovers characteristic oscillating shapes, the doping
dependence of the phase shift and also special features which we
discussed in detail in previous sections, such as particle-hole
(Fano like) asymmetries and spin-charge separation.  We note that
the LL parameters $K$ in our case are usually quite close to one,
and as a result the signature of the interactions is not
pronounced. We also note that, at finite temperature (or
frequency), one does not strictly reach the ``open boundary fixed
point''; this point should be especially emphasized in our case,
where some of the backscattering $V_{b\alpha}$ are weakly
relevant.

Apart from the ``standard'' propagators that were introduced
above, there is also an additional part connected to the band
mixing
\begin{eqnarray}
\fl
\langle\psi_{L\pi}^{\dag}(x,t)\psi_{Ro}(x,0)\rangle=r_{o\pi}\exp\imath
    ((k_{Fo}+k_{F\pi})x+\delta_{o\pi})\prod_{\nu=c,s}F_{\nu}(x,t;\sqrt{V_{F\pi}V_{Fo}},\sqrt{K_{\pi}K_{o}}) \nonumber \\
    +\exp-\imath ((k_{Fo}+k_{F\pi})x+\delta_{o\pi})\prod_{\nu=c,s}F_{\nu}(-x,t;\sqrt{V_{F\pi}V_{Fo}},\sqrt{K_{\pi}K_{o}})\label{corr pi}
\end{eqnarray}

The last component determines the difference between the chains
caused by an impurity. This gives rise to a new zig-zag pattern in
the vicinity of the impurity. The contribution to the LSW coming
from this part has a different periodicity
($(k_{Fo}+k_{F\pi})^{-1}$) along the ladder than the intraband
components and it alternates in the y direction so it should be
detectable by Fourier transform of an STM spectrum. As was pointed
out in the previous section, band mixing favors $\pi$-CDW
instanton tunelling and the existence of this extra part in
$A(\omega , x)$ is in full agreement with the physical argument
presented there.

\begin{figure}
  \centerline{\includegraphics[width=\figwidth]{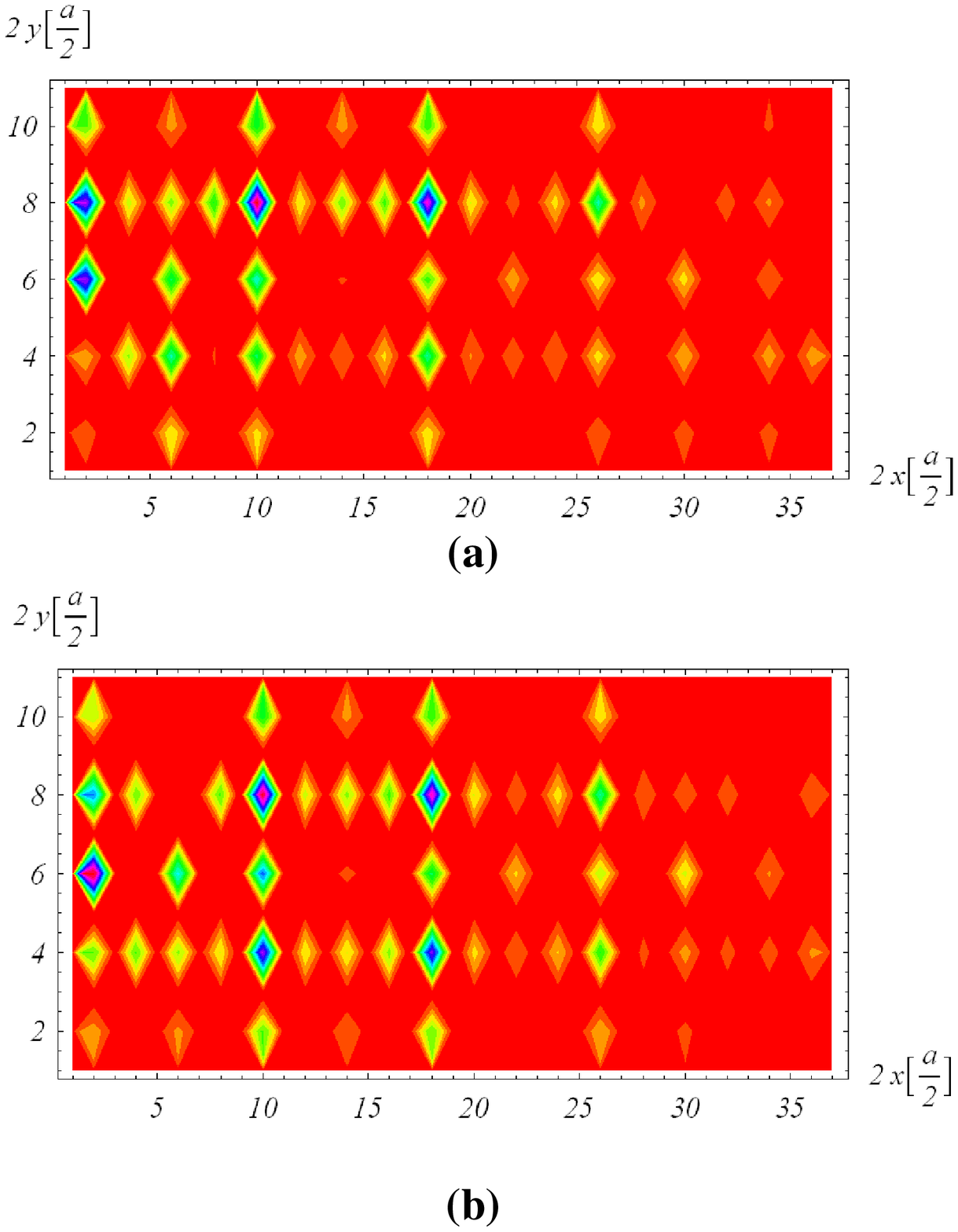}}
  \caption{Alternating part of the LDOS around an impurity. The impurity sits in the cell at the left of the left boundary of each panel (hence is not shown). Interference patterns around (a) a strongly asymmetric impurity (Zn substitution of a $Cu$ atom for example) and (b) a weakly asymmetric one (for example on-rung or apical $O$); $\beta=1/T=100$; the values of the $K_{i}$ parameters are the same as in Table \ref{tab:refl}}\label{fig:profile}
\end{figure}

These effects are illustrated on the density plots shown in
Fig.\ref{fig:profile}. In order to obtain a realistic plot, it
is important to use the correct relative amplitudes for the
various components. For that, one needs to know the values of
$r_{oo},~r_{\pi\pi},~r_{o\pi}$ at some given temperature, for a
given type of defect within the unit cell. We calculated these
coefficients, using the method presented in the Appendix and we
give their values in  Table \ref{tab:refl}, for
$T=0.1\Lambda_{0}$   (which, assuming $\Lambda_{0}\sim t$,
corresponds to $100K$) and for various positions of the defect
in the unit cell.
\begin{table}
  \centering
  \begin{tabular}{|c|c|c|c|c|}
    \hline
    / & Cu C2S2 & O central C2S2 & Cu C2S1 & O central C2S1 \\
    \hline
    $r_{o}$ & 0.17 & 0.71 & 0.14 & 0.65 \\
    $r_{\pi}$ & 0.63 & 0 & 0.21 & 0 \\
    $r_{o\pi}$ & 0.26 & 0 & 0.19 & 0 \\
    \hline
  \end{tabular}
  \caption{Reflection coefficients due to defects in the \emph{C2S2} and \emph{C2S1} phases; these are calculated using the method described in \ref{TMatrix} (for the two types of substitution); The values of the $K_{i}$ parameters are
  $K_{s\pi}=K_{so}=1$, $K_{co}=0.9$, $K_{c\pi}=0.95$ and the inverse temperature is $\beta=1/T=10$; the reflection coefficients increase with decreasing temperature}\label{tab:refl}
\end{table}
The reflection coefficient for the on-leg oxygen atom is very
similar to that of the copper atom, except for the fact that
the interband reflection coefficient $r_{o\pi}$ is reduced. The
amplitudes given in the table are calculated at temperatures
such that the LL picture holds, and that interactions do not
play a crucial role. Decreasing the energy scale will
renormalize these to higher values.

The most straightforward way to determine the imaginary part of
the Green function is to probe the local density of states
(LDOS), using STM. Two types of measurements are available:
constant voltage scan of the topography and  I-V
characteristics at a given atom location. The former technique
is commonly used to image surface electron waves interference
patterns. Using Fourier transform and filtering, one is able to
extract  the alternating part of LDOS which contains important
information. The most interesting one in our case is the
difference between profiles obtained for symmetric and
asymmetric impurities. This is shown in  Fig.\ref{fig:profile},
where the real space, alternating, part of the LSW is shown in
the contour map (as in experiments); $x$ and $y$  denote the
on-leg and on-rung  coordinates, respectively
One may clearly recognize the expected exponential decay of the
amplitude, and the interference effects caused by three
components with different periodicities; the pattern generated
by on-leg oxygen sites is not identical to that due to on-rung
oxygens, because of different decompositions into the band
components. Because of the interference effect, the shape of
the pattern will change with doping (the wavelengths are doping
dependent). Fig.\ref{fig:profile} also shows  that the effects
induced by the impurity are mostly confined within a single
ladder (the amplitudes on the outer oxygens are rather small).
We find that, as the temperature decreases, the pattern has
an increasing spatial extension and that the position of its
maximum shifts away from the impurity; its shape may also
changes, because the relative amplitudes will renormalize.
Interaction effects are not pronounced,
but they control this renormalization.

Two features are worth pointing out; one is the relatively
large amplitude predicted on the on-rung oxygen orbital,
proving that it is important to include this atom in the
calculation. This effect will be quite strong whenever $r_{o}$
is large. The other is the presence of a zig-zag pattern, which
is particularly visible for the $Cu$ atoms which are close
enough to the impurity. It shows that this kind of DW is locked
on the impurity, even if band mixing is weak. This confirms the
qualitative prediction that we made for the tunelling regime.
The amplitude of the zig-zag pattern will increase upon
decreasing the temperature. If we  take into account the fact
that, by and large, asymmetric impurities reside primarily on
the ladder,  while symmetric ones are caused by sources
external to the ladder, one is given an effective tool to
recognize the nature of the impurity. Comparing bottom and top
panels in Fig.\ref{fig:profile} can thus help one identify
``on-ladder'' and ``out-of-ladder'' impurities. In our case, we
find that in the bottom panel, their effect is more symmetric
while the asymmetry between the legs is more pronounced in the
top panel.

The latter technique in an STM experiment probes the frequency
dependence of the LSW at a given site. The characteristic
oscillations and particle-hole asymmetry are the effects caused
by the impurity.
\begin{figure}
  \centerline{\includegraphics[width=\figwidth]{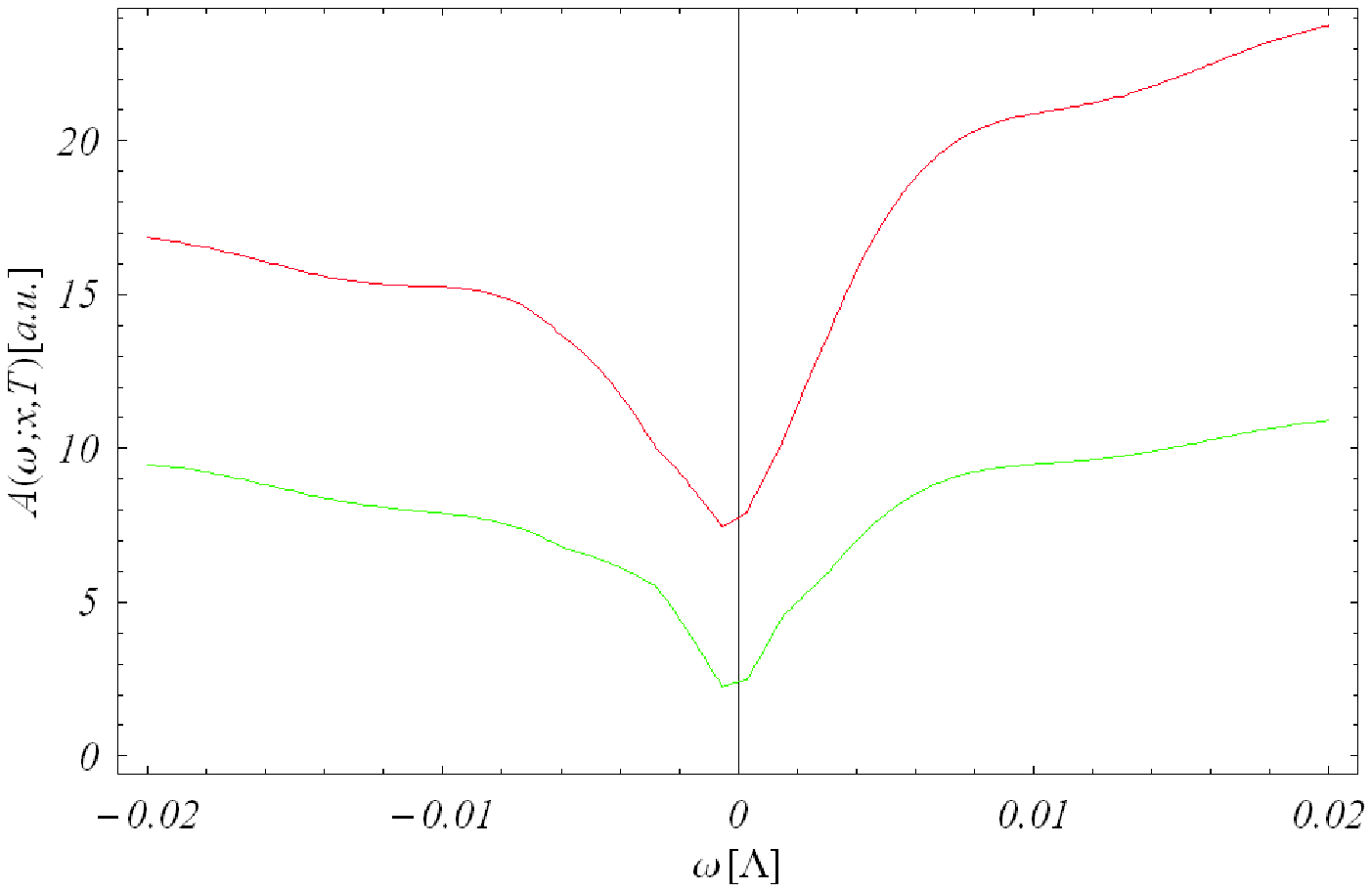}}
  \caption{Frequency dependence of the LDOS measured on copper atoms (the case of non-symmetric impurity);
  the distance from the impurity is $x=2$, in lattice units. Red pertains to
  atoms located on the same leg as that where the impurity sits, green to
  atoms on the the leg  opposite to where the impurity sits.}\label{fig:LSWfreq}
\end{figure}
Fig.~\ref{fig:LSWfreq} refers to the gapless situation; the
minimum amplitude is obtained close to zero frequency and
then it increases. For the gapped phases this feature will be
shifted to frequencies of order of a gap value. This effect
could in principle allow one to detect the gaps if they are
present in the system.


\subsection{Knight shifts}

As we showed in the previous section, the main effect caused by an
impurity is the mixing of left and right moving carriers. This is due to
the fact that, in the vicinity of the impurity, the wave
vector k is not a good quantum number anymore. One can thus
expect that the mixing of $0$ and $2k_{F}$ wave functions caused by
an impurity will also affect other quantities.

Let us focus on the magnetization profile given by the local, zero
frequency susceptibility. Assuming that we are subjecting the
sample to some external, uniform magnetic field, the $0$ and
$2k_{F}$ parts will generate alternating magnetization patterns. This
effect can be detected by the changes of Knight shift for the
atoms in the vicinity of the defect. Only the spin part
contributes here. The uniform part does not contain band mixing
terms, so we expect only intraband contributions. The expectation
value that we are interested in, is
$\langle\psi_{R\alpha}^{\dag}\psi_{L\alpha}\rho_{\alpha}\rangle=\langle\exp(\imath
2k_{F\alpha})\exp(-\imath\phi_{R\alpha}+\imath\phi_{L\alpha})\frac{d\phi_{\alpha}}{dx}\rangle$.
This quantity is known for the single spin chain case
 \cite{Eggert_gen1}, so in our case the finite temperature solution for a band ($o$ or $\pi$)
is given by the following expression (valid in the spin-rotational invariant case, where $K_{s}=1$):
\begin{equation}\label{eq:susc}
    \chi_{alt}(x)\simeq\int_{-1}^{1}du\frac{\sqrt{v\beta\sinh(\frac{2\pi x}{v\beta})}}{u^{2}+1-(u^{2}-1)\cosh(\frac{2\pi x}{v\beta})}
\end{equation}
where x is the distance from the impurity. Eq. (\ref{eq:susc})
describes a spin mode LL oscillation with a $2k_{F}$ wave
vector around the impurity. It produces a characteristic
envelope, decreasing as $\sqrt{x}$ when $|x|\rightarrow 0 $,
and also decreasing, but exponentially, when $|x|\rightarrow
\infty $, with a maximum amplitude of the order of a few
lattice spacings for temperatures on the order of $100$K. This
shape was obtained previously in the case of a spin chain
\cite{Eggert2}, and we will call it \emph{wing-like}, in the
following.

\begin{figure}
  \centerline{\includegraphics[width=\figwidth]{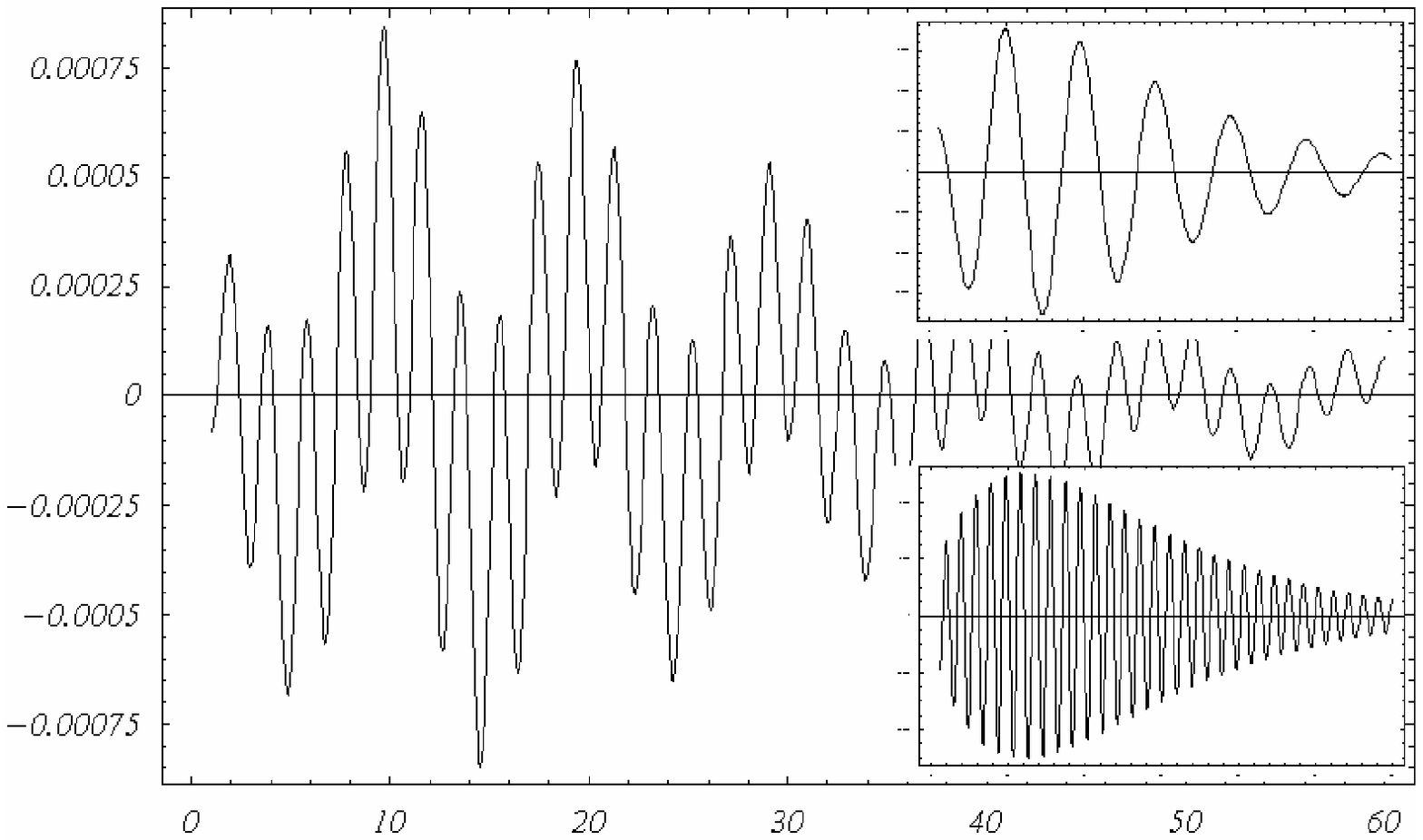}}
  \caption{Interference pattern due to two intraband alternating
  magnetization profiles shown in the insets (magnetization,
  given in arbitrary units, as a function of a distance from impurity,
  in lattice cell units); the figure shows the amplitudes at the locations of $Cu$ nuclei.
  These nuclei feel the spatially varying magnetic field is which proportional to the interference pattern. The
  doping is $\delta=0.3$, and the inverse temperature is $\beta=200$}\label{fig:Knight1}
\end{figure}

We already pointed out \cite{kakashvili_LSW} that the important
effect caused by backscattering, in the two-band case with
different wave vectors $k_{Fi}$, is the interference - the
modulation of alternating wings. For the static susceptibility
the effect in the two chains will be the same, the reason being
that the auxiliary magnetic field used as a probe is uniform
and hence the same for the two chains. An example of two
intraband alternating magnetizations (inset) and the resulting
pattern when they are added is shown in Fig.~\ref{fig:Knight1}
(there is no band mixing component here). As the temperature is
lowered, in the case of a single chain, characteristic peaks
appear on opposite sides of the main line (the reason is that
for the lowest temperatures many atoms feel similar relatively
strong additional fields). In the case of ladders, interference
effects suppress the peaks and the distribution of local fields
is more homogenous.
In the above  formulae the limit of the ``open boundary fixed point''
was considered. In reality, there will some coefficient in front
of it describing the intraband reflection. In the case of
``out-of-ladder'' impurities (symmetric case) the reflection in the
$o$ band is much stronger than in the $\pi$ band, so that interference
effects should be much less pronounced, and the broadening is quite
similar to that of the single band problem.

At large distances we find an exponential decay of the alternating
part, with a localization length
$\eta_{loc}=V_{\alpha}/2\pi T$. Two conclusions can be drawn from
this; one is that the broadening of the Knight shift line has a
characteristic temperature dependence (the lower the temperature,
the larger the number of nuclei  affected by the alternating
magnetization; this number increases exponentially with $1/T$
enhancing the amplitude or the broadened part). This broadening is
different from the usual thermal broadening or from the
temperature independent broadening due to the internal currents,
in an orbital current phase (OCP). The other is that $\eta$
is different for the two bands (the $V_{\alpha}$ influence) and
one has always $\eta_{o}>\eta_{\pi}$; this effect is strongly
pronounced as one approaches the bottom of the $\pi$ band. The
conclusion is that with increasing doping the influence of the
$\pi$ band decreases.

As we have already mentioned, Eq. (\ref{eq:susc}) describes how
the magnetization is affected by the presence of an impurity.
One of the quantities which is measured in an NMR
experiment is the strength of the interaction between a nuclear spin and
its electronic environment, which is usually dominated by the contact
Pauli interaction between a nuclear spin and s-electrons on a
given atom. To second order in perturbation, one can construct virtual hoppings
of band carriers onto this core s-orbital. This process explains how
uniform LL susceptibilities affect the observed NMR lines. By the
same argument, we find that NMR lines (or more precisely their
broadening) will be affected by the alternating magnetization
induced in the band. From an experimental perspective, let us
point out that, in the presence of band interferences (see
Fig.\ref{fig:Knight1} ), the shape of the broadening is difficult to
predict and to distinguish from the other broadening mechanisms (in
contradistinction with the single chain case).

\begin{figure}
  \centerline{\includegraphics[width=\figwidth]{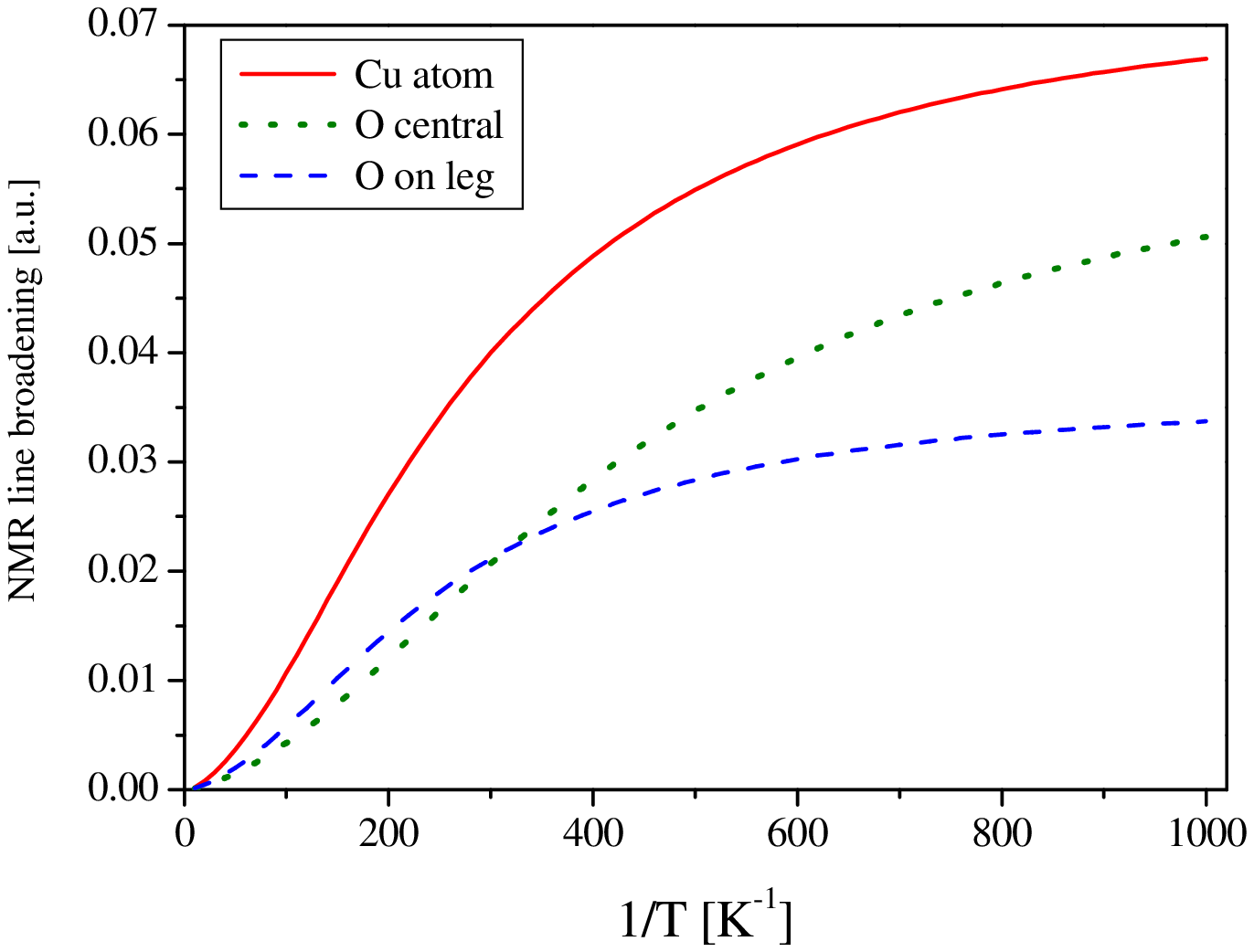}}
  \caption{The integrated line broadenings on different atoms
  given as a function of temperature.}\label{fig:Knight_atom}
\end{figure}

To determine the broadening, we need to compute the hopping
amplitudes between the core s- and the p,d- electronic orbitals on
neighboring atoms. This was done perturbatively starting from a
defect-free ladder giving us access to the broadenings for
different atoms, shown in Fig.\ref{fig:Knight_atom}.
Their main characteristics are as follows.
The NMR line
broadening of the on-rung oxygen atom (which couples only to the
$o$ band, so there is no interference) is the only one to display the
characteristic shape found in the case of a single chain.
The temperature dependence of the broadening
is similar for all the nuclei. One can thus single out this type of broadening, by
performing measurements at different temperatures.
Only the linear (intermediate) part of each curve is generic.
Different atoms will indeed give different slopes for the curves, since
the two bands contribute differently to the wave function on a
given atom . One has to remember that the large $x$ part of
$\chi(x)$ decays exponentially with a characteristic length
proportional to the Fermi velocity in a given band.
These band interferences will disappear with increasing doping,
because the velocity in the $\pi$ band becomes extremely small and
only the influence of the $o$ band remains. This effect and the
different values of  $\lambda$ (Eq. (\ref{bazy1})) cause
different slopes, but the first effect is quite easy to extract in
our formalism. What remains is just the amplitude of the
broadening which will change from atom to atom, providing us with
information about the high momentum component of a given orbital,
in the LL.

A few words about the saturation effect at very low temperatures
are in order. For one single impurity, the $1/T$ temperature
dependence should persist down to zero temperatures, implying that
the integrated line broadening goes to infinity when $T\rightarrow
0$. But this situation is unphysical: one needs a finite number of
impurities to detect any signal. What is more one can expect some
gaps in the system (like a spin gap in the \emph{C2S1} phase, or
some collective disorder effects), so there is always some
low-energy cut-off. In order to account for this, we introduce an
upper limit in the integral ($x_{max}=200$ in the example above,
which corresponds to the typical impurity concentration
($n=0.005$) used in experiments). If the impurity concentration
gives the low-energy cut-off scale (assuming that the Fermi
velocities are known) then, from this low temperature saturation
value, one can gain some insight into the density of impurities in
the sample. Logarithmic corrections due to boundary operators can
also be evaluated \cite{brunel_log, Affleck_boundarymarginal}.
They are stronger than the contributions from bulk operators
($\ln^{4}T$ instead of $\ln^{0.5}T$ for $K=1$).  One might have
hoped to detect these at the lowest temperatures -- presumably
then in the \emph{C2S2} phase. But even in that case, marginal
operators are very weak in the entire critical phase, and it is
not even clear that they can be detected in experiments, so we
will not pursue in this direction and we omit these logarithmic
correction from hereon.

Let us briefly discuss correlation functions for two $2k_{F}$
operators (four fermion correlations). They are needed to
obtain the NMR relaxation rate $T_{1}^{-1}$. Such correlations,
including charge phase fields with $K_{c}\neq1$ were evaluated
in Refs.~ \cite{brunel_log, Wakita_boundary_T1}. In our case,
as for the Green functions, $oooo$, $\pi\pi\pi\pi$ and
$oo\pi\pi$ terms will be present (the others are forbidden by
perpendicular momentum conservation) and will produce
interference patterns. This calculation (showing the behavior
in the vicinity of an impurity) was done analytically for the
case of two modes in a LL \cite{Wakita_boundary_T1}. In our
case, at high dopings, the eigenbasis is $B_{o\pi}$ and it
gives us precisely two modes ($o$ and $\pi$). So we can extend
the results of Ref.~\cite{Wakita_boundary_T1}  to our problem.
The result is similar to that for the static Knight shift: the
lower the temperature the more atoms are affected. The problem
is that the spatial fluctuation of $T_{1}^{-1}$ is more
difficult to access experimentally (especially when $K_{i}\sim
1$), and  its temperature dependence is not so easy to analyze.
There are some additional peaks with $4k_{F\alpha}$ periodicity
but their amplitude is proportional to the interaction strength
(namely $1-K_{c+}$) so, in our case, this effect is too weak to
be detected. Hence we will not consider $T_{1}^{-1}$ any
further, in the remainder of this paper.

\section{Discussion} \label{discussion}

\subsection{Experimental relevance}

The above calculated alternating part of the susceptibility can be
detected experimentally  as an additional broadening of the NMR
peaks, since the local magnetic field felt by each atom (which is
proportional to the resonance frequency) will depend on its
distance to the impurity.

From an experimental perspective, in order for our predictions to
be testable, we need to specify the regimes where our free LL
analysis applies. A short (theoretical) answer would be: as long
as the impurity is strong enough a perturbation such that there
exists an energy window where the results of BCFT (LL regime) can
be used. In the low temperature limit, this requirement might not
be fulfilled. In the \emph{C2S2} phase, higher order effects and
phase stability in the vicinity of impurity might take us away
from the LL state; for the \emph{C2S1} phase, the existence of a
gap in the $o$ band spin mode is a limitation. Last but not least,
rather than a single impurity, a sample generally contains a
finite concentration of defects. If these cannot be treated as
independent scatterers, one expects collective effects to take
place. We address these issues below.

\subsection{Defect in the low doping \emph{C1S0} phase}\label{sec:C1S0}

The situation is very different at low doping, where the
disorder-free ladder is in the \emph{C1S0} phase.
This particular limit was briefly mentioned when we discussed
the RG procedure for the backscattering potential. A strong
chain asymmetry emerges during the RG flow and an unpaired spin
appears near the impurity. The physics, in this case is that of
the Kondo problem. Here, we summarize previous attempts to
describe this regime and we give some arguments which allow us
to make experimentally testable predictions.

Close to half filling, and in the defect-free case, we found a
phase in which all the LL modes are gapped at zero temperature,
except one, corresponding to the motion of a charge soliton
along the ladder. The fluctuations of the gapped modes are
frozen, thus one may assume that they will adjust to a boundary
(described by an unknown fixed point). The existence of this
new, intermediate fixed point of the RG flow was already
revealed in a previous discussion. The appearance and the
on-chain nature of the boundary state was suggested. In our
case, we note that, in the low temperature, unperturbed state,
we have a strong, on-rung coupling. Then, we expect that an
asymmetric impurity leaves one spin unpaired, such that there
should exist a boundary state with spin one-half. These kind of
states and their magnetic properties have already been
discussed in the literature in DMRG studies
 \cite{Laukamp_DMRG_imp, Lauchli_Li_imp}, but also analytically
using refermionization methods \cite{Lecheminant_Majorana} (one
works then in the space of singlet and triplet Majorana
fermions). The lowest lying excitation (inside the gap) is the
triplet, which corresponds to an overscreened $1/2$ spin. The
magnetization profiles as well as other experimentally
accessible properties of the boundary state were discussed in
detail in Ref.~\cite{Lecheminant_Majorana}

What about the gapless $c+$ mode? Spin-charge separation still
holds when the impurity is present, so the above spin state
will not couple to free holons at $T=0$: the only free soliton
moving along the semi-infinite ladder is the holon in the $c+$
charge mode. Fermionic tunelling events are forbidden by the
fact that the lowest excitations are only of bosonic type, and,
as we already pointed out, higher order tunelling processes are
not likely because all $K_{i}\sim 1$. The problem reduces to
the reflection of a soliton on a barrier, which was  studied in
the context of a quasi-classical sine-Gordon model
\cite{schuricht_gapped_strong_V}; since, for the other modes,
the low energy physics is dominated by gaps their solution
describing single soliton scattering should remain valid.

We now turn to finite temperatures. In the \emph{C1S0} phase, the
LL modes are defined in the $B_{+-}$ basis so that the band mixing
(backscattering) effect caused by the impurity should not affect
the values of the gaps. For simplicity we assume that the two spin
gaps are equal. The relative magnitudes of the gaps and of the
impurity potential vary with doping, but we may safely assume that
upon reaching the temperature corresponding to the smaller (spin)
gap (corresponding to the bosonic $s+,~s-$ modes or singlet and
triplet Majorana fermion masses), reflection is still relevant. In
other words, the new fixed point, emerging because of the spin-state
induced on the impurity, still dominates the physics when $T\sim \Delta_{s}$.

For temperatures below $\Delta_{s}$, carriers from the bulk
cannot screen the impurity spin (spin degrees of freedom are
frozen). It was then suggested that the impurity can be
screened by boundary states which arise from the Luther-Emery
model (from a Majorana fermion picture, when both masses are
positive). In that case, screening will depend on the occupancy
of this boundary state pinned at the Fermi energy. Whether or
not such screening does occur depends on the relative sizes of
the energies $J_{K}$ and $\Delta_{s}$: if the latter is much
smaller, we expect an interband Kondo regime in the SCd phase.
This is a slight modification of the model with intraband
$J_{K}$ described in detail by Le Hur \cite{Le_Hur_imp}. This
model yields non-standard temperature dependences of the
correlation functions (the screening boundary state will arise
through some sort of ``proximity effect''). Recall however
that, for our model,
 all the $J_{K}$ are strongly irrelevant in this regime,
so that one is likely to obtain a Curie-like behavior (uncoupled
impurity spin) as the temperature goes to zero.

At higher temperatures ($T>\Delta_{s}$),
spin degrees of freedom can fluctuate and $J_{K}^{o\pi}$
dominates the physics. We expect Kondo physics, as described by Eq. (\ref{eq:Kondo_eff}), to be relevant.
 This defines an interband Kondo
problem with finite transmission of spinons through the impurity.
Refermionization of both bulk and
impurity spin degrees of freedom allows us to show that the physics in this regime is that of a
resonant level model with interactions (see Appendix \ref{Kondo} for details).

At still higher temperatures $T \gg \Delta_{s+}$ , the
backscattering strength $V_{o\pi}$ becomes weaker than the
largest charge transverse gap. In such a case, which appears
most likely, in view of our RG analysis, the impurity physics
is totally masked by the gaps and the high temperature
crossover regime (around $\Delta_{c-}$) will be described by
the weak backscattering potential for the soliton
\cite{tsvelik_gapped_weak_V}.

In the opposite case, when the impurity potential is stronger
than the gaps (this can arise for very strong initial values of
reflection coefficients), there exists a high temperature
regime where, in principle, we should recover the results that
we established in the previous sections (but this time
$V_{o\pi}$ is the dominant backscattering). Starting from the
high temperature limit, we have free fermions, and, as the
temperature is decreased, first we encounter an interband
tunelling regime through the barrier (still with free fermions
in the bulk). As $T$ is reduced further, the $c-,~s-$ gaps
open, singlets form on the rungs and a spin state appears on
the boundary. The crossover, intermediate regime, would be
described by the logarithmic corrections to the free LL
\cite{brunel_log}. The point is that this regime will appear at
temperatures of order $1000K$, so it is beyond the
experimentally accessible range.

\subsection{Defect in the intermediate doping \emph{C2S2} phase: stability in the vicinity of the impurity}

The most difficult case to analyze is the \emph{C2S2} phase,
since it is critical.
Following the lines of our analysis for the low and high doping
regimes, we would study how an impurity behaves in the \emph{C2S2}
LL.  However, this might turn out to be an ill defined problem.
The reason is because this critical phase is extremely sensitive
to perturbations. In the case of a non-magnetic impurity in a
conventional metal, Friedel oscillations are generated in the
vicinity of the defect. Here, a similar effect could occur,
leading to some charge/spin instability. In this case the defect
could induce some ordering in its vicinity and then interact with this
new environment.

It is known that a finite concentration of impurities is a
relevant perturbation for many 1D systems and in the next section
we will explicitly show that this is the case for the $Cu$-$O$
ladder as well. This type of distortion can change the phase diagram, i.e the
thermodynamic states. By contrast a single impurity, no matter how
strong it is, cannot modify the bulk properties of the ladder.
However we may ask whether it could lead to an instability of the
phase in its neighborhood.
Unfortunately, we are in a situation where interactions and
impurity backscattering are competing and momentum is not a good
quantum number anymore. So, to answer the question, using a full
RG analysis would require the introduction of momentum dependent
vertices.

A similar question was raised in the context of a single chain
problem \cite{Andergassen_fRG} where the flow of the
self-energy was obtained in a functional RG (fRG) formalism.
For the flow of bulk operators, one performs RG both in
momentum and frequency space, whereas for the point impurity
operator, one integrates out all spatial dependence and one is
left with a frequency space RG; the situation close to the
boundary is somewhere in between these two limits and this is
the reason why layer by layer real-space RG is necessary. There
are far more couplings in the ladder problem than in the chain
problem (sixteen as opposed to four), so we were not able to
perform a full fRG procedure. However, under the reasonable
assumption that vertex corrections do not change dramatically
during the flow, we were able to connect the standard RG and
fRG. Investigating possible differences enables us to make some
predictions for the case of the $Cu$-$O$ ladder.

Two factors affect the RG differential equations in the
vicinity of a defect. One plays a role near the lower quantum
critical point (QCP), on the \emph{C2S2} side of the
\emph{C2S2}-\emph{C1S0} border. It is the band mixing
reflection which accounts for the change of the diagonal,
non-interacting, basis of reflected carriers as was described
in detail in section \ref{sec:bound}. In other words, band
mixing introduces a diagonal basis for free carriers that
differs from the $B_{o\pi}$ bands. This mixing reduces the
difference between Fermi velocities in the two
\emph{eigen}-bands, so it will change the initial
$P_{i},~Q_{i}$ parameters in the RG equations; namely, for a
given doping, the $Q_{i}$ parameters will be smaller (and these
coefficients decrease like $\sim r_{o\pi}^{2}$). This effect
may be also understood if we remember that the mixing term
generates an asymmetry between the chains; it can be taken into
account by imposing an artificial difference between the
chemical potentials, which was shown \cite{wessel_inhom_doping}
to change the ratio between $V_{Fo}$ and $V_{F\pi}$.
This effect tends to shift the phase boundaries towards higher doping.

The other factor affects the upper QCP, on the \emph{C2S2}
side of the \emph{C2S1}-\emph{C2S2} border. This effect is
slightly more subtle, and it is responsible for shifting the
\emph{C2S1} boundary towards smaller dopings. In a previous
section, Sec. \ref{sec:Green}, we showed that electron propagators
change significantly in the vicinity of the impurity. So we
proceed in following way: free electron Green functions enter the
RG equations when, for each energy scale $\Lambda$, we integrate
out particle-particle and particle-hole instabilities
$I(\omega)|_{\Lambda}$ (four fermion correlators). In the bulk
case these two terms are proportional and their difference can be
absorbed through the introduction of coefficients
$\tilde{\alpha}=\frac{V_{Fo}}{V_{Fo}+V_{F\pi}}$. When an impurity
is present  this simplification does not work anymore. The
asymmetry between the two bands (and thus also between the Cooper
and Peierls channels) will be enhanced by the following term
\begin{equation}\label{as1}
    I_{P}-I_{C}=r_{oo}^{2}\int I(-\omega ;~V_{Fo})d\omega -r_{\pi\pi}^{2}\int I(\omega ;~V_{F\pi})d\omega
\end{equation}
where $I(\omega)$ contains an additional asymmetry caused  by
Fano-type, interference effects, as pointed out in
Ref.~\cite{kakashvili_LSW}.

The $g_{o}$ term, which is responsible for the DW gap opening in
the highly doped phase, becomes even more relevant, since the
mechanism that we described in the previous paragraph tends to
increase the band asymmetry. This is the reason why the intraband
DW phase (which dominates at large dopings) is more rigid. This can
be understood as a direct consequence of the fact that the density
wave, which is present in the ground state of the \emph{C2S1}
phase, can adjust its phase, and hence the charge distribution
near the defect, in order to minimize the energy.

We evaluated numerically  the values of both (Cooper- and Peierls-
type) contributions and we found that they are of the same order.
We also checked, using Table. \ref{tab:refl}, that for most of the
positions of the impurity within the elementary cell,
$r_{oo}^{2}-r_{\pi\pi}^{2}\sim ~r_{o\pi}^{2}$. Thus, we cannot
tell here which of the intra- or inter- band instability dominates,
but we can make clear statements about the rigidity of the intermediate gapless phase.

Both contributions described above tend to suppress this phase
and to open some gaps in the spectrum. Two additional arguments
support this scenario. One is that that any marginal operator
will affect correlation functions much more strongly close to a
boundary. This comes from the change in dimensionality of
Greens functions and it was predicted \cite{brunel_log} to give
$\log(r)^{2}$ instead of $\log(r)^{1/2}$ corrections to $G(r)$.

A computation of $2k_{F}$ instabilities, such as
$\langle\cos(\phi(x))\cos(\phi(0))\rangle$ \cite{brunel_log}
shows that they are enhanced or suppressed close to the
boundary depending on
whether the bulk LL parameters $K$ are smaller or larger than one 
The analytic form of
these instabilities strongly suggests that the orderings are not
only favored close to the boundary but that they are also
contributing to the real space RG calculations (in the same way as
the fermionic Green functions $I_{P/C}$ introduced above). Based
on the above considerations (taking into account the bulk values
of $K_{i}[\Lambda]$), we find that, on the large doping side, a DW
inside the o-band is enhanced in comparison to the other
instabilities, while on the small doping side
$\theta_{1},~\theta_{3}$  are enhanced and $\phi_{2}$ is
suppressed. The last field describes the $\pi$-CDW instability. This
remark is fully consistent with our discussion of the tunelling
regime; this instability corresponds to the instantons at small
dopings ($m_{o\pi}$ dominates) and at large dopings ($r_{o}$
dominates) respectively. This suggests that the results of our first order RG
calculation should still hold when higher order terms are taken
into account.



\begin{figure}
  \centerline{\includegraphics[width=\figwidth]{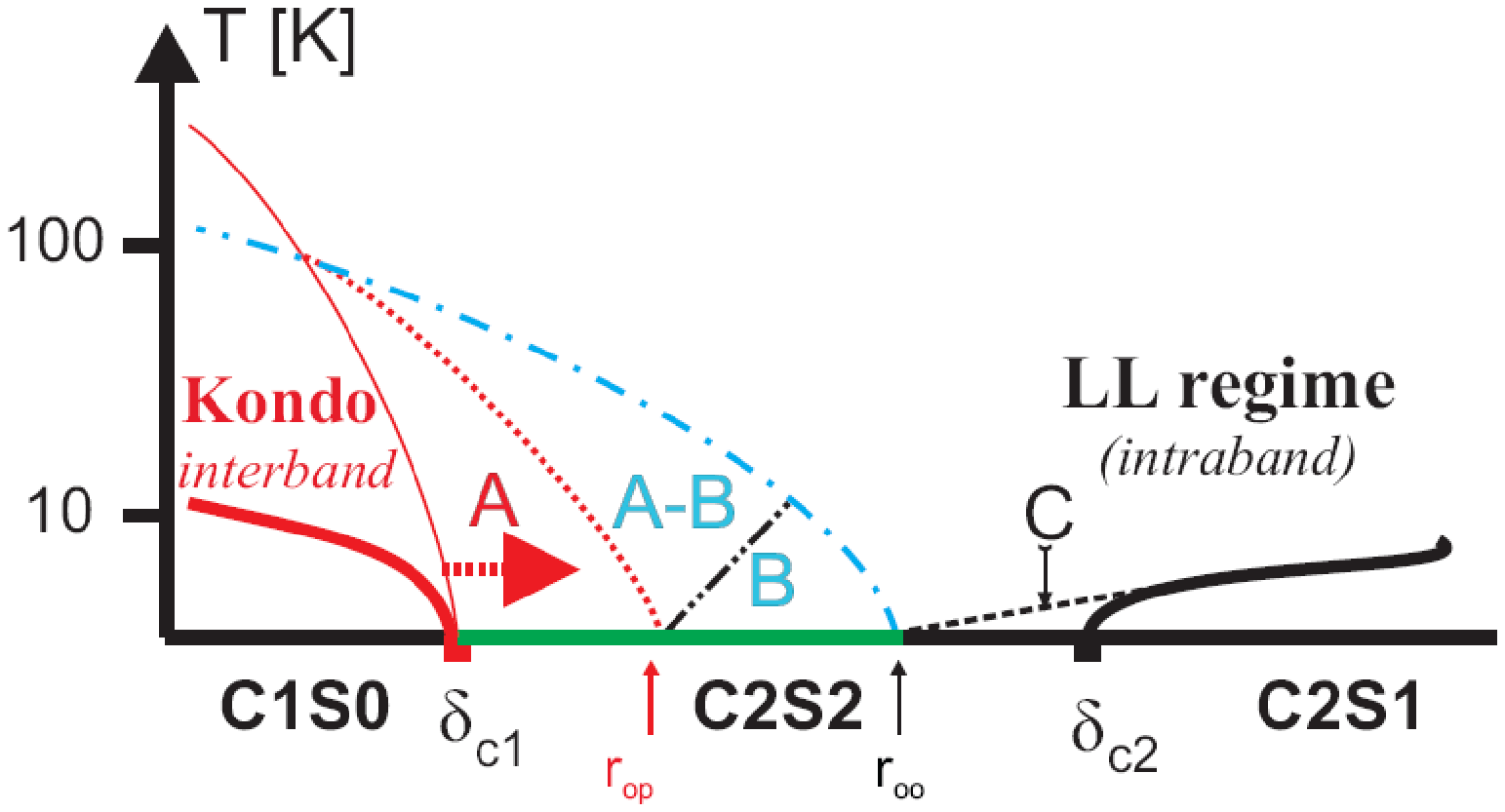}}
  \caption{Influence of a non-magnetic impurity on the \emph{C2S2} region of the two-leg ladder
phase diagram. Solid lines mark the temperatures below which bulk gaps develop. Dashed, dotted and
dashed-dotted lines mark the cross-over temperatures of the various screening regimes observed near the
impurity.
 A spin 1/2 is present in the impurity unit cell in the region below the blue dashed-dotted line. We distinguish the regions of interband Kondo physics (A) and single channel ($o-$band) Kondo physics (B) (the latter results from the channel $J_{K,o/\pi}$ asymmetry). The key observation is the transition from interband (red color) to intraband (black color) physics; the nature of the Kondo effect is different on both sides (however we predict finite, zero temperature, transmission in both cases -- thick green line). On the higher doping side there exists a region (C) of spinless $o$-band DW instability induced by the intraband backscattering on the impurity \footnote{This is the most general phase diagram, the details of which will depend on microscopic structure of obstacle: in the case $V_{o\pi}=0$ (highly unstable situation) one would expect C region only}; region A arises if $K_{c-}$ becomes larger than one (the condition of $J_{o\pi} relevance)$); region B will appear only if neither $V_{o\pi}$ nor $V_{o,\pi}$ are enough strong to dominate the physics.}\label{fig:diag_C2S2}
\end{figure}

The following picture emerges from our stability discussion of the
\emph{C2S2} critical gapless phase: it is suppressed both on the
small and large dopings sides. On the low doping side it is caused
by additional band mixing on the obstacle; on the high doping side
it is the intraband instability which adjusts well to the presence
of the impurity (and can even be enhanced in comparison with the
bulk). The phase diagram that pertains to the \emph{C2S2} phase is shown in
Fig.\ref{fig:diag_C2S2}.

The high doping side is simple -- we can use a similar description
to that we would give for the \emph{C2S1} phase (the dominance of
the intraband LL physics with a vanishing gap in the $\phi_{so}$ mode). The spin sector remains in the
total/transverse basis, but there is no gap in the $c-$ mode,
which was a critical requirement in order to get equal couplings
for the two channels and to make $J_{K}^{o\pi}$ more relevant than
the other Kondo couplings. In addition, intraband fluctuations in
the charge sector are present, which also tends to differentiate
the bands. As a result, interband couplings do not dominate anymore,
and a new relevant (intraband) Kondo coupling $\Delta
J_{K}=J_{K}^{o}-J_{K}^{\pi}$ appears. The two-channel Kondo
(intraband) physics is unstable with respect to band asymmetry;
instead we simply expect a single channel coupling with the spin from each
half-infinite ladder. Because of the presence of backscattering,
single-channel Kondo physics should describe this regime. The physical
nature of the screening is far more traditional than in the \emph{C1S0}
regime described by Eq. (\ref{eq:Kondo_eff}). Intraband
physics can dominate even if there exists an unpaired spin-state
dynamically induced on the impurity, because the BCFT framework
still describes this regime, even if it is not  perfectly
reflecting anymore. To be more specific, if this spin-state
emerges, the marginally relevant instability is described by the Kondo
problem, as for the critical
phase of the single chain \cite{Le_Hur_imp, Eggert_RGcomp}, 
with some finite transmission through the impurity.

The low doping side it is also dominated by the Kondo exchange
interaction, but then one expects
$\pi$-DW fluctuations close to the impurity, because of the relevant
interband interactions. As a result one would expect a similar
spin screening description to that we have given for the high
doping side, but emerging in the interband regime. Because of
the interband $r_{o\pi}$ scattering, the low doping regime ($B_{+-}$
basis) is present around the impurity, so $J_{K}^{o\pi}$ should be
still enhanced during the RG flow.
 The low energy physics
is then likely described by the interband Kondo terms of Eq. (\ref{eq:Kondo_eff}). 
Near the upper and lower doping boundaries of the \emph{C2S2} phase we predict finite spinon
transmission through the impurity. Nevertheless, as discussed above, the physics is quite different
at these two ends.
Thus, there should be a phase transition somewhere in the
\emph{C2S2} phase, driven by the competition between
$J_{K}^{o\pi}$ and $J_{K}^{o}$. The microscopic structure of the
defect, which determines the initial conditions of RG, becomes
relevant, so it is difficult to predict the position and the
properties of this phase transition. We leave this topic for
further studies.


\subsection{ Influence of the spin gap in the \emph{C2S1} phase:}

In the high doping, \emph{C2S1}, regime the situation is quite
clear; $V_{o\pi}$(or $t_{o\pi}$) is only marginal
 and the low energy physics favors the $B_{o\pi}$
basis (this is garanteed by the existence of an intraband gap).
Above the temperature scale of the gap, even in the strongly
asymmetric case, the density of states on the impurity is much
higher in energy than the Fermi points, which precludes the
formation of boundary states. The open boundary fixed point is
thus quite robust. Some arguments suggest that this result
could be extrapolated, to a certain extent, below the gap. The
$o$ and $\pi$ bands are well separated then and the gap affects
only the spin sector of the $o$ band. Let us focus on this
particular band (it is clear that for the $\pi$ band the LL
picture will hold down to zero temperature). The density wave
which forms inside this band tends to get pinned on the
impurity.

The opening of a gap does not significantly affect the physics
around the impurity. The arguments given before, as well as
results of numerical studies, rule out the possibility of a new
boundary state on the impurity. From numerical studies of spin
ladders, it appears that these kinds of states can appear only
if the on-rung exchange energy is very different from the
on-leg exchange energy ($J_{\perp}\ll J$), or in the presence
of an additional strong magnetic field on the boundary. This is
definitely not the case in the \emph{C2S1} phase. This
statement is also in agreement with an analysis based on the
boundary sine-Gordon model with $K_{co}\sim 1$ (no
spinon-antiholon bound states)
\cite{schuricht_gapped_strong_V}. A picture of interfering LL
modes backscattered on non-magnetic boundary emerges from this
analysis.

When the scattering due to the impurity becomes strong, it simply
favors a given configuration of the LL phase fields. If we assume that
a gap opens, it freezes some configuration of solitons
within the $\phi_{so}$ mode. The opening a gap can be seen as the slowing
down of this mode: the expresionss that were derived before are still
applicable but with renormalized velocities.
Beyond this scaling limit, for the lowest temperatures one can
use the solution found by Tsvelik et al.
\cite{tsvelik_gapped_weak_V}. Using spin-charge separation we
find that the charge mode in the $o$ band still gives rise to a
wing-like shape of propagators, whereas the spin sector has an
activated behaviour described by Bessel function of the second
kind with no spatial dependence (exactly like in the bulk
gapped case). As a result, the shape of the LDOS should be
similar to the form predicted in the previous section, except
for the fact that the  maximal amplitude is shifted to a
frequency $\omega\sim \Delta$. The NMR line broadening should
saturate at the lowest temperatures (there are two competing
terms with exponential temperature dependences).

Despite the presence of a spin gap in both the \emph{C1S0} and the
\emph{C2S1} phases, the physics is very different in the spin
sector for the two phases. In particular, there does not appear to
be a bound spin state around a defect in the \emph{C2S1} case, but
rather a DW instability, as a result of spinon interferences. The
origin of this effect comes from the presence of very different
dominant boundary operators in the two states.

\subsection{The many impurity case}

Until now the discussion was restricted to the problem of a single
impurity and to the physical effects induced in its vicinity. In
this last part, we turn to the problem of a finite concentration
of weak scatterers to find the effects caused by impurities acting
collectively.

As for the single impurity problem, we will start with a discussion of
forward scattering, which can be treated in a similar fashion to
the single impurity case, using the proper shifts of the phase fields
on each impurity. 
As for the disordered single chain, the shifts of the
fields $\phi$ are responsible for example for the exponential
decay of some density-density correlations involving $\phi$ fields
(it will not affect the current operator though, which depends
solely on $\theta$ fields, and so this effect does not correspond
to weak localization). To illustrate this effect, let us take some
order operator expressed as a functional of bosonic phase fields,
for example in the case of interband DW:
\begin{equation}\label{eq:forw_dis1}
    O_{A}(r)=F[\phi_{c+}, \phi_{s+}, \theta_{c-}, \theta_{s-}]
\end{equation}
we find an additional exponential decay in the form
\begin{equation}\label{eq:forw_dis2}
  O_{A}(r)\sim \exp(-\frac{2K_{c+,s+}^{2}D_{f}r}{V_{c+,s+}})\exp(-\frac{2K_{c-,s-}^{-2}D_{f}r}{V_{c-,s-}})
\end{equation}

where $D_{i}=nV_{i}^{2}$ (as usual for gaussian, uncorrelated disorder with $V_{i}$ potential on each local impurity).


The characteristic localization lengths are expected to be
different for different operators and their computation is
straightforward. Let us only remark, that the above
formalism applies so long as a given mode is described in the LL
framework (i.e is effectively massless). In the case when an unpaired
spin is induced on the impurity (Kondo regime in the low doping case)
one should also take into account an additional phase shift of the
$\phi_{s+}$ field; it is the only mode which can induce impurity spin
flips, which corresponds to a shift of the boundary condition.

There is one caveat. Phase
shifts affect the initial conditions of the RG flow, both for the
interactions and for backward scattering. As we pointed out, in
the case of a single, strong impurity,  backward and forward processes are
independent; hence,
inter-
and intra- band terms may be affected in a different way, leading, for instance, to different
localization lengths, from Eq. (\ref{eq:forw_dis2}). This may modify
the initial conditions of the backscattering contribution to the
RG flow. Nevertheless, as long as we consider short range, Hubbard-type
interactions, this effect can be safely neglected.

Including the backscattering part, we investigate Anderson
localization -- the instability of the phase diagram of the
clean system to collective disorder. This problem has been
discussed  in the context of single orbital two-leg ladders
\cite{Orignac_disorder} by means of a simplified RG approach.
The physically relevant defect-free phases appear to be very
fragile with respect to weak localization effects. An RG
analysis \cite{Fujimoto_disorder} to second order in
perturbation, for the half filled case, confirms this
statement: in order to preserve the stability of the Mott
state, one needs to choose initial values of the interaction
terms several orders of magnitude larger than their disorder
counterpart. Otherwise collective disorder will easily prevail
during the RG flow and one observes the switchover from Mott to
Anderson transition.
 In our case, the RG
procedure follows closely that developed in
Ref.~\cite{Fujimoto_disorder} (see Appendix
\ref{collectivedisorder}), so we only summarize the results
without going into computational details.

For the $Cu$-$O$ ladder, we find that disorder terms are
strongly relevant and dominate the RG flow. One new
feature is revealed by our analysis; it is related to the basis
rotation caused by disorder terms. One finds two generic flows
in the presence of disorder:
\begin{itemize}
    \item when intraband disorder dominates we have
    Anderson localization for the charge modes; all interaction
    terms are irrelevant except $g_{1}$ and $ g_{2}$ which produce a gap in the
   $o$ and $\pi$ spin modes; this favors an intraband  CDW order, similar to the chain problem
    \item when interband disorder dominates, it causes
        Anderson localization and interaction terms are
        irrelevant, except $g_{1a}$, which favors a
        $\pi-CDW$  type order and produces a $o/\pi$ mixing
        down to the bottom of the bands
\end{itemize}

As for the discussion of interaction effects, interband
instabilities are likely to dominate at lower dopings; the driving
parameter is the ratio $D_{intra}/D_{inter}$ at the beginning of
the flow. One has to remember, that, until a characteristic energy scale is reached
during the RG procedure, impurities
renormalize as single, independent entities. In such a case the first part of
the flow will determine the initial ratio
$D_{intra}/D_{inter}[\Lambda_{c}]$ at a scale $\Lambda_{c}\sim
n_{imp}^{-1}$.




Finally, two extra remarks  about the interplay between the
collective and  single strong impurity limits are in order.

First, collective disorder competes with interactions (making
these less relevant) so, in some indirect way, it would tend to
suppress the opening of gaps at lower temperatures.  For example, in
the \emph{C2S1} phase, the low energy physics is dominated either
by disorder or by the spin gap, and in the former case
intraband solitonic states arise on each impurity (a hierarchy of
breathers appears when disorder
renormalizes $K_{i}$ to zero).
These issues were addressed in some numerical studies, but their
detailed discussion is beyond the scope of the present work.

Second, as for the single impurity problem, we observe distinct
small and large doping regimes. Collective and single impurity
effects reinforce each other and the regions of the phase
diagram where inter- and intra- band physics dominate seem to
coincide for the two cases. For the high doping regime, at
intermediate energy scales (where the screening clouds starts
to overlap), one may expect some crossover behavior between the
two power law behaviors characterized by either a single
impurity or the disorder exponent. At low doping, the situation
is not so clear, because of an additional flow of
$K_{c-}[l]\rightarrow 0$ induced by disorder. Thus, the Kondo
regime should be suppressed, but a detailed discussion of this
effect remains an open question.

\section{Conclusions}

Two main results were obtained in this paper. First, correlation
functions for the doped, two-leg, $Cu$-$O$ Hubbard ladder in the
vicinity of a single, strong, non-magnetic impurity were evaluated
in the regime where the system is in a gapless LL state. In this
case, these correlators can be given in closed analytical form.
This means that we were able to generalize the method used to
treat the equivalent problem for a single chain, to construct
chiral fields mappings, to handle band mixing on the impurity.
Using these correlation functions, we were able to compute various
magnetic response functions, which are relevant experimentally in
NMR spectroscopy.
We computed expressions for the local spectral response functions,
which allowed us to determine the density of states versus energy
on different sites. This local spectral weight is accessible in STM measurements,
which makes our predictions testable. We have emphasized the
different signatures  of ``on-ladder'' and ``out-of-ladder''
impurities. The conclusion is that one can
distinguish these two kind of impurities experimentally.

Second, we discussed and analyzed the low and high doping regimes
where, at low temperature, gaps are present. We found that the impurity physics, in the low and high doping phases, is very different.
The local stability of the critical \emph{C2S2} phase was also examined
and we were led to conclude
that the low energy physics (not protected by any gaps) is severely affected by impurities. In this respect, it is important to recall that this phase only appears because we have included oxygen atoms in the structure, and this implies that different bases describe the charge and spin fluctuations.  We examined all possible scenarios for the \emph{C2S2} phase and we find that finite spin transmission takes place in some part of this doping range. This effect shows the strong relevance of oxygen atoms in the description of ladder compounds.

Finally, we considered the
evolution of the phase diagram of the $Cu$-$O$ ladder when a
finite concentration of weak scatterers is present. Localization
effects that had been described for chains and for single orbital
ladders are found here too. However, near the
\emph{C1S0}-\emph{C2S2} boundary, we find that new instabilities
which are subdominant in the bulk may develop.
\vspace{1cm}

\noindent {\bf ACKNOWLEDMENTS}

\vspace{0.5cm}
This work was supported in part by the Swiss NSF under MaNEP and Division II and by an ESRT Marie Curie fellowship.

\appendix

\section{Interband Kondo fixed point}\label{Kondo}

The Kondo problem for a LL has been extensively discussed in
the literature so, in this section, we only highlight the
specifics of our model. The Hamiltonian for the intraband part
is known, for example from Ref.~\cite{Le_Hur_imp}, in the
bosonic field language
\begin{equation}\label{eq:K_intra}
\fl    H_{b-imp}^{\alpha\alpha}= (J_{z}-4v_{F})S_{z}\nabla\phi_{Rs+}(x=0)
+J_{xy}[S_{+}\cos(\phi_{Rs-}(x=0))+h.c.]
\end{equation}
and this corresponds to a two-channel Kondo model.

For the interband part we find

\begin{equation}\label{eq:K_inter}
\fl   H_{b-imp}^{\alpha\bar{\alpha}}=
    \cos(\phi_{Rc-}(x=0))(J_{z}S_{z}\cos(\phi_{Rs-}(x=0))
+J_{xy}[S_{+}\exp(\imath\phi_{Rs+}(x=0))+h.c.])
\end{equation}

The charge transverse mode dominates the RG equations. In
order to evaluate the corresponding terms, we have to bear in mind that we are working
with OBC, so that the scaling dimension of each chiral field is
$K^{-1}$.
Assuming that we are already at the fixed point
basis $B_{+-}$ (a more detailed treatment shows that this
simplification is justified), we get for interband effects
\begin{equation}\label{es:K_RG1}
    \frac{dJ_{z}}{dl}=(2-(K_{c-}^{-1}[l]+K_{s-}^{-1}[l]))
\end{equation}
where the functional $K_{i}[l]$ is known from the bulk RG flow.
From previous studies of two-leg ladders
\cite{chudzinski_ladder_rapid,chudzinski_ladder_long} we have
that $K_{c-}\gg 1$ and  $K_{s+/-}\geq 1$ at the beginning of
the flow, but $K_{s+/-}\rightarrow 0$ at the end. From these
observations we may infer the strong relevance of interband
terms in comparison with the others.

It is also possible to refermionize our low energy spin
Hamiltonian (\ref{eq:K_inter}) using Kivelson's
transformation: $S_{+}\rightarrow U d^{\dag}$ (where the Klein factor
U takes care of Fermi statistics). This transformation to an
interacting resonant level model enables us to extract some information regarding the
thermodynamics. The result is

\begin{equation}\label{eq:K_refer}
\fl   H_{b-imp}=J_{z}Re[\psi_{Rs-}](x=0)(d^{\dag}d-1/2)\\
+J_{xy}[d^{\dag}\psi_{Rs+}(x=0)+h.c.]
\end{equation}
Let us point out that two different fermionic operators are
responsible for, respectively, interactions (first term) and hopping (second
term) onto the resonant level. The interaction term does not have the usual
density form.

\section{System of RG equations for the disordered
case}\label{collectivedisorder}

The form of the backscattering potential is the same for the
single impurity and for the collective disorder cases. The only
difference is the physical meaning of certain quantities; for
the single, strong impurity case, $V$ is simply the strength of
the $2k_{F}$ component of the impurity potential; for the many
weak impurity case, $V$ is replaced by $D$, the density of
impurities in the system, multiplied by the single impurity
strength (this quantity is assumed to be non-divergent in the
scaling limit). In real space it is a random variable with a
Gaussian distribution, and it affects the form of the RG
equations: in the former case there is no scaling in real space
(in time only), whereas in the latter case one may consider
that disorder spreads over the entire system. Because of this,
collective disorder is able to renormalize the interactions
$g_{i}$ and the LL \emph{K} parameters, and also to generate
additional non-linear terms. All these effects were taken into
account previously \cite{Fujimoto_disorder}. Our task at this
point, is simply to generalize the procedure, accounting for
the basis rotation, and to check whether the conclusions of
Ref.~\cite{Fujimoto_disorder} remain valid in the case of our
doped $Cu$-$O$ ladder. To that end, we introduce intraband
terms $D_{o}$ and $D_{\pi}$ (denoted $D_{intra}$) and interband
terms (denoted $D_{inter}$). We assume that
 these quantities are non-zero and statistically independent.

As for the disorder-free system, when oxygen atoms are included, one needs
to introduce angles of rotation of the eigenbasis with respect
to the $B_{+-}$ basis, through the quantities
 $P_{i} (\neq 1)$ and $Q_{i} (\neq 0)$ \cite{chudzinski_ladder_long}
($i=s,c$ for the charge or spin sector).
 We start from the system of RG equations derived in
Ref.~\cite{Fujimoto_disorder} and make the following
substitutions
\begin{eqnarray}
K_{s-}\rightarrow P_{s}^{2}K_{s-}+Q_{s}^{2}K_{s+}\\
K_{s+}\rightarrow P_{s}^{2}K_{s+}+Q_{s}^{2}K_{s-}\\
K_{c-}\rightarrow P_{c}^{2}K_{c-}+Q_{c}^{2}K_{c+}\\
K_{c+}\rightarrow P_{c}^{2}K_{c+}+Q_{c}^{2}K_{c-}
\end{eqnarray}
For the interband term $D_{o\pi}$ (defined in the $B_{+-}$
basis), and all the other ``gauge'' terms, this task is
straighforward and the procedure does not require any further
comments; as for the intraband terms, we rewrite them as
$D_{o+\pi}=\frac{1}{2}(D_{o} + D_{\pi})$ corresponding to the
operator $\cos\phi_{s-}\cos\phi_{s+}\cos\phi_{c-}\cos\phi_{c+}$
and $D_{o-\pi}=\frac{1}{2}(D_{o}-D_{\pi})$ corresponding to
$\sin\phi_{s-}\sin\phi_{s+}\sin\phi_{c-}\sin\phi_{c+}$ which
now can be treated in much the same way as the $g_{o \pm \pi}$
terms of the pure case \cite{chudzinski_ladder_long} (these two
were noted as $g_1$ and $g_2$ in that paper). The two RG
differential equations describing the flow of these sum and
difference of intraband backscatterings will be then slightly
different from those given in Ref.~\cite{Fujimoto_disorder}. We
have instead
\begin{eqnarray}
\fl \frac{dD_{o+\pi}}{dl} = D_{o+\pi}\cdot (3-\frac{1}{2}(K_{c+}+K_{c-}+K_{s+}+K_{s-})) \nonumber \\
+P_{c}Q_{c}(K_{c+}-K_{c-})P_{s}Q_{s}(K_{s+}-K_{s-})D_{o-\pi}
\end{eqnarray}

\begin{eqnarray}
\fl \frac{dD_{o-\pi}}{dl} = -D_{o-\pi}\cdot (3-\frac{1}{2}(K_{c+}+K_{c-}+K_{s+}+K_{s-})) \nonumber \\
+P_{c}Q_{c}(K_{c+}-K_{c-})P_{s}Q_{s}(K_{s+}-K_{s-})D_{o+\pi}
\end{eqnarray}

In addition we also need to take into account the influence of
disorder terms on the RG flows describing the bases rotations.
Using the convention $P_{s}=\cos\alpha$ and $P_{c}=\cos\beta$,
we have
\begin{equation}\label{eq:alpha}
\fl  \frac{d\cot2\alpha(\beta)}{dl}=
\frac{((dK_{s-(c-)}-dK_{s+(c+)})\tan4\alpha(\beta)+
dB_{s-s+(c-c+)})}{K_{s-(c-)}-K_{s+(c+)}}\cdot dl^{-1}
\end{equation}
where the disorder terms cause changes both in the diagonal (standard)
and in the non-diagonal parts $B_{s-s+(c-c+)}$ of Luttinger liquid compressibility matrix. The change, with respect to clean case,
 of the non-diagonal element (mixing $i-th$ and $j-th$ modes) is:
\begin{equation}\label{eq:change_diag}
\Delta \frac{dB_{ij}}{dl}= P_{i}Q_{i}(D_{o\pi}^{2}(1-K_{i}K_{j}))-K_{i}K_{j}h[P_{i}]D_{o-\pi}D_{o+\pi}
\end{equation}

The functional $h[P_{i}]$ is given by:
\begin{equation}
 h[P_{i}]=((P_{i}Q_{i})^{2}+0.25(P_{i}^{2}-Q_{i}^{2}))^{-1}
\end{equation}

\section{Calculation of backscattering amplitudes}\label{TMatrix}


In order to determine the properties of the $Cu$-$O$ ladder in the vicinity
of an impurity
we need to determine the transmission and reflection coefficients
of the barrier (i.e of the unit cell containing the defect). Their
values depend on the position of the defect within the unit cell --
which contains two $Cu$ and five $O$ atoms.

We wish to evaluate transmissions in all the channels (both intra-
and inter- bands) at some finite temperature; we take the cell
containing the impurity and connect it to two semi-infinite leads
(ladders) described by four LL modes with arbitrary
compressibility $K_{i}$. We had to assume arbitrary values,
because we use $K_{i}[\Lambda_{T}]$, where the renormalization of
$K_{i}(l)$ is stopped at the energy scale $\Lambda_{T}$
corresponding to T. We have to work at finite temperature, because
we evaluate backscattering potentials $V_{i}[\Lambda_{0}]$ (the
initial conditions of RG flow) which then will flow towards a low
energy regime ($\Lambda\rightarrow\infty$).

We follow a  method which is widely used in the context of
mesoscopic systems. It allows one to solve 1D transmission problems
and the results closely match those obtained by means of functional RG techniques
 \cite{agarwal_r_NEGF}. We consider the Hamiltonian of the
perturbed part and then introduce couplings to the leads via
appropriate self energies. The retarded Green function of the
\emph{i-th} elementary cell ($i$ is the distance from the defect)
is obtained using Wick's equation (the bare Green functions are
perturbed by hopping onto the semi-infinite leads described by the
surface Green functions)
\begin{equation}\label{eq:1}
    G_{i,i}=(E-H_{i}-\Sigma_{L}-\Sigma_{R})^{-1} ,
\end{equation}
The Hamiltonian of the perturbed part, $H_{i}$, is explicitly
introduced in this formalism, which enables us to modify it
freely and to place the impurity at any chosen location in the unit
cell. Self energies can be obtained from the surface Green
functions $ g_{i,i}^{L/R} $
\begin{equation}\label{eq:2}
    \Sigma_{L/R}=T_{i,i\mp1}\cdot g_{i\mp1,i\mp1}^{L/R}\cdot
    T_{i\mp1,i}.
\end{equation}
For the surface Green functions we take the functions derived
by Eggert for the LL with OBC \cite{kakashvili_LSW}. This
allows us to include interaction effects quite easily, through
changes of the \emph{K} parameters; the massive mode case can
also be treated using, for instance, results obtained by
Tsvelik \cite{tsvelik_gapped_weak_V}. The limitation of this
method is that it uses a linear dispersion approximation (low
energy limit).

The rest of the calculation is done using standard non-equilibrium
Green's function formalism, under the coherent transport
assumption.  In the linear response
regime, the formula describing the flow of charge through the
system reduces to the Landauer-Buttiker expression. The
transmission function (the key quantity of this formalism
corresponding to the conductivity) is:
\begin{equation}\label{eq:4}
    T(E)=Tr(\Gamma_{L}(E)\cdot G(E)\cdot \Gamma_{R}(E)\cdot
    G^{\dagger}(E)),
\end{equation}
where $ \Gamma_{L/R} $ is
\begin{equation}\label{eq:6}
    \Gamma_{L/R}=\imath\cdot(\Sigma_{L/R}-\Sigma_{L/R}^{\dagger}).
\end{equation}

\bibliographystyle{prsty}
\bibliography{totphys,ladder3b,impurities}

\end{document}